\documentstyle[11pt,aaspp4]{article} 
\newcommand{\kms}{km~s$^{-1}$}

\newcommand{\subsun}{\mbox{$_{\odot}$}}
\newcommand{\etal}{{\it et al.\/}}
\newcommand{\teff}{$T_{\mbox{\scriptsize eff}}$}
\newcommand{\grav}{log($g$)}
\newcommand{\diff}{$\Delta$[Fe/H]$_{\rm II-I}$}

\lefthead{Carretta {\it et al.} }
\righthead{Extremely Metal-Poor Stars}

\begin{document}
\title{Stellar Archaeology: a Keck Pilot Program on Extremely Metal-Poor 
Stars From the Hamburg/ESO Survey. II. Abundance Analysis\altaffilmark{1}}

\author{Eugenio Carretta\altaffilmark{2},
Raffaele Gratton\altaffilmark{2}, Judith G. Cohen\altaffilmark{3},
Timothy C. Beers\altaffilmark{4}, \& Norbert Christlieb\altaffilmark{5}}

\altaffiltext{1}{Based in large part on observations obtained at the
        W.M. Keck Observatory, which is operated jointly by the California 
        Institute of Technology, the University of California
        and NASA,}
\altaffiltext{2}{INAF - Osservatorio Astronomico di Padova, Vicolo dell'Osservatorio 5,
        35122, Padova, Italy}
\altaffiltext{3}{Palomar Observatory, Mail Stop 105-24,
        California Institute of Technology, Pasadena, CA \, 91125}
\altaffiltext{4}{Department of Physics and Astronomy, Michigan State University,
East Lansing, Michigan 48824-1116}
\altaffiltext{5}{Hamburger Sternwarte, Gojenbergsweg 112, D-21029, Hamburg, 
Germany}

\begin{abstract}

We present a detailed abundance analysis of eight stars selected as extremely
metal-poor candidates from the Hamburg/ESO Survey (HES).  For comparison, we
have also analysed three extremely metal-poor candidates from the HK survey,
and three additional bright metal-poor stars.  With this work, we have {\it
doubled} the number of extremely metal-poor stars ([Fe/H] $\le 3.0$) with
high-precision abundance analyses.  Based on this analysis, our sample of
extremely metal-poor candidates from the Hamburg/ESO survey contains three
stars with [Fe/H] $\le -3.0$, three more with [Fe/H] $\le -2.8$, and two
stars that are only slightly more metal rich.  Thus, the chain of procedures
that led to the selection of these stars from the HES successfully provides a
high fraction of extremely metal-poor stars.

We verify that our choices for stellar parameters, derived in Paper I, and
independent of the high-dispersion spectroscopic analysis, lead to
acceptable ionization and excitation balances for Fe.  Substantial non-LTE
effects in Fe appear to be ruled out by the above agreement,
even at these extremely low metallicities.

For the $\alpha-$elements Mg, Si, Ca, Ti, the light element Al, the iron-peak
elements Sc, Cr, Mn, and the neutron-capture elements Sr and Ba, we find trends
in abundance ratios [X/Fe] similar to those found by previous investigations.
These trends appear to be identical for giants and for dwarfs.  However, the
scatter in most of these ratios, even at [Fe/H] $\le -3.0$ dex, is surprisingly
small.  Only Sr and Ba, among the elements we examined, show scatter larger
than the expected errors.  Future work (the 0Z project) will provide much
stronger constraints on the scatter (or lack thereof) in elemental abundances
for a substantially greater number of stars.

We discuss the implications of these results for the early chemical evolution
of the Galaxy, including such issues as the number of contributing SN, and the
sizes of typical protogalactic fragments in which they were born.

In addition, we have identified a very metal-poor star in our sample that
appears to represent the result of the s-process chain, operating in a very
metal-poor environment, and exhibiting extremely enhanced C, Ba, and Pb, and
somewhat enhanced Sr.

\end{abstract}

\keywords{Galaxy: halo --- stars: abundances --- Galaxy, evolution}

\section{INTRODUCTION}

Stellar archaeology is the study of present stellar generations in order to
infer the characteristics of a previous stellar generation that no longer
exists. This is one of the primary aims for investigations of the chemical
composition of extremely metal-poor Population II stars, as they provide
important clues to the properties (e.g., mass, composition) of the very first
objects formed in the Galaxy, the so-called Population III stars.

Long-lived, slowly evolving main-sequence dwarfs are quite suitable for this
purpose, since they retain in their atmospheres the elements produced by
previously born massive stars that exploded as Type-II supernovae (SN).
Unlike the stars presently in the giant-branch stage of evolution, main-
sequence stars are expected to be unaffected by internal mixing during their
lifetime (although, in some cases, they may exhibit the spectral signatures of
contamination from  material transfered from close, evolved, companions).

Here we adopt the definition given in Cohen \etal\ (2001; hereafter Paper I),
and we consider only stars with [Fe/H] $\le -3$ to be extremely metal poor
(hereafter EMP).\footnote{We use the usual spectroscopic notation: log~n(A) is
  the abundance (by number) of the element A in the usual scale where
  log~n(H)=12, while [A/H] denotes the logarithmic ratio of the abundances of
  elements A and H in the star, minus the same quantity in the Sun.}

This definition seems almost straightforward, since all previous investigations
(see the review by McWilliam 1997) revealed that at [Fe/H] = $-2.4$ many
elemental ratios [X/Fe] (where X is Ba, Sr, Cr, Al, or Mn) display a sudden
change in the slope of the relationship of [X/Fe] versus [Fe/H]. McWilliam
\etal\ (1995; hereafter McW95), and others since, have suggested that the
patterns observed at very low metallicity can be explained by assuming that a
stochastic mechanism is at work, with only a few supernovae responsible for the
observed enrichment patterns.  By selecting those stars with [Fe/H] $\le -3$
we can be certain that we are sampling a regime where stars were polluted by
the very first supernovae, in an environment likely to have been rather
different from that in which the bulk of Galactic stars formed.

The literature concerning EMP stars is continously increasing, as ever more
efficient spectrographs at large telescopes come on-line, analysis
techniques are refined, and laboratory measurements of fundamental atomic
parameters required for detailed abundance analyses are carried out.  Interest
in this area of astrophysics arises for a number of reasons, since the study
of these objects provides insights into such relevant issues as the early
chemical evolution of the Galaxy, nucleosynthesis by zero-metallicity massive
stars, and the role of the {\it r}-process and the {\it s}-process in building
up the presently observed abundances of neutron-capture elements in stars.
Theories of the nucleosynthesis mechanisms themselves benefit from direct
comparison with observed abundance ratios in EMP stars, in order to tune model
yields. On the other hand, by use of the predicted yields from SN of different
masses, one might attempt to decode the observed run of abundances as a
function of metallicity, in order to derive the range of numbers and masses of
SN contributing to the chemical enrichment in various environments, as well as
the epochs of the building up of the Galactic elements (see, e.g.,  Karlsson \&
Gustafsson 2001).

Moreover, a direct link to the distant Universe is provided by dating methods
that use cosmo-chronology (age estimates based on the radioactive decay of
unstable heavy nuclei in EMP stars) to provide independent measurements of the
ages of the oldest stars in our Galaxy (Sneden \etal\ 2000, Cayrel \etal\
2001a,b, Toenjes \etal\ 2001), which can be compared to the ages of other
apparently primordial objects such as globular clusters, derived by different
methods (see Carretta \etal\ 2000 and references therein).

The shortcoming, up to now, has been the small size of available stellar
samples, due to the relative rarity of EMP stars, their faint apparent
magnitudes, and the need for high-resolution, high signal-to-noise (hereafter,
S/N)  spectroscopy to derive their elemental abundances with
suitable precision.
The presently available sample of such stars is simply too small
for statistical studies that might provide strong constraints on the
above-mentioned problems. In fact, summing up all previous high-dispersion
analyses of very metal-poor stars (those with [Fe/H] $\le -2.0$), the total 
sample with published detailed analyses hardly reaches 50 objects.

We are mainly interested in studying the mechanisms involved in the early
chemical evolution of the Galaxy.  The large intrinsic spread in (some)
elemental ratios found at extremely low metallicities requires a very large
database, not only to properly sample the observed trends, but more
importantly, to quantify the scatter in the observed elemental abundance
distributions as a function of declining metallicity.  An increase
of available sample sizes for EMP stars by an {\it order of magnitude} is
required to fully understand the nature of the very first generations of
Galactic stars.

In our ongoing study, we intend to exploit the recently completed Hamburg/ESO
Survey (HES; Wisotzki \etal\ 1996, Christlieb \etal\ 2001a,b) to significantly
increase the number of EMP candidates with available high S/N,
high-dispersion spectroscopy. Herein we present the results of the Keck Pilot
Program on EMP stars, in which we test the ability of the HES to deliver a
large sample of newly identified EMP stars for abundance analysis.

The selection, observations, and data reduction of the present sample are
discussed at length in Paper I; the present paper will deal only with the
abundance analysis. In \S 2 there is a brief summary of relevant information
given in Paper I. The equivalent width measurements and
tests of their quality are described in \S 3.
The derived abundances are presented in \S 4, and
discussed in \S 5. The last section summarizes our present results.

\section{OBSERVATIONS, DATA REDUCTION, AND ATMOSPHERIC PARAMETERS}

The selection, observational details, and the preliminary data reduction of our
program stars are discussed in Paper I. Here we briefly summarize the essential
information.

Eight candidate EMP stars from the HES were observed with the HIRES
spectrograph (Vogt \etal\ 1994) 
at Keck I on two nights in Semptember 2000.  On the same
nights we also acquired spectra for three EMP candidates from the HK survey
(Beers, Preston \& Shectman 1985, 1992),
as well as three well-studied bright metal-poor stars as comparisons.
One of the HES stars turned out to be a re-discovery of a star from the HK
survey (HE~2344$-$2800 = CS~22966-045). The relevant parameters of the
observations, as well as photometry and the adopted data analysis for all stars
in our sample are given in Table~1 of Paper I.

A spectral resolving power of $R = 45,000$ was used with an 0.86 arcsec slit
projecting to 3 pixels in the HIRES focal plane CCD detector, resulting in
spectra covering the region from 3870 to 5400\,{\AA}, with essentially no gaps.
The figure of merit, F, as defined by Norris, Ryan \& Beers (2001; NRB2001), is
$\sim 600$ for the {\it worst} of our spectra, which guarantees the high
quality of our observational material.  In this paper, we have {\it doubled}
the sample of high precision (F $> 600$) spectra available for stars with
[Fe/H] $<-3.0$, including three newly discovered EMP stars.

Some examples of the spectra are shown in Paper I.

\subsection{Adopted Atmospheric Parameters}

The procedure used to derive \teff\ estimates for our program stars is fully
explained in \S 4-6 of Paper I. Very briefly, \teff\ is derived from
broad-band colors, taking the mean estimates deduced from the de-reddened $V-K$
and $V-J$ colors. We used the grid of predicted broad-band colors and
bolometric corrections of Houdashelt, Bell \& Sweigart (2000), based on the
MARCS stellar atmosphere code (Gustafsson \etal\ 1975), and corrected the
colors for reddening by adoption of the extinction maps of Schlegel,
Finkbeiner \& Davis (1998) (see Table~1 in Paper I).

With \teff\ fixed, the gravity \grav\ was obtained using the Y$^2$ isochrones
(Yi \etal\ 2001); we adopted the 14 Gyr, [Fe/H] = $-3.3$ isochrone.  For the
star HD 140283 we adopted the \grav\ obtained by Korn \& Gehren (2000), derived
from the Hipparcos parallax.

Holding \teff\ and \grav\ fixed, the final overall metallicities [A/H] for the
stars were obtained iteratively, by matching observed equivalent widths (EWs)
with the synthetic ones computed by integrating the equation of transport at
different wavelenghts along each line for the flux, extracted from a model
atmosphere in the grid of Kurucz (1995), with no 
overshooting\footnote{Models are interpolated linearly in \teff\ and 
logarithmically in the other quantities. Note
that model atmospheres for stars with [Fe/H]$< -3$ are not interpolated, but
{\it extrapolated}, since the grid of Kurucz does not have models below this
metallicity.}. In fact, Castelli, Gratton \& Kurucz (1997) noted that Kurucz 
models with the convective overshooting option switched off better reproduce 
observables in stars other than the Sun. 
Microturbulent velocities $v_t$ were derived by eliminating any
trend in derived abundances of Fe I lines with the expected EWs (see Magain
1984).

The adopted atmospheric parameters are summarized in Table~\ref{t:atm}.

\section{EQUIVALENT WIDTHS}

Equivalent widths were measured from the one-dimensional, normalized spectra
using an automatic routine that determines a local continuum level for each
line by an iterative clipping procedure. A fraction $C$ of the 200 spectral
pixels centered on the line to be measured is used; the highest 200$C$ pixels
are the initial dataset for this process. After various tests, we adopted
$C=1$ for the spectra of all stars, except the three very bright stars and the
two giants from the HK survey. For these stars $C=0.5$ was used.

\begin{figure}
\epsscale{1.0}
\plotone{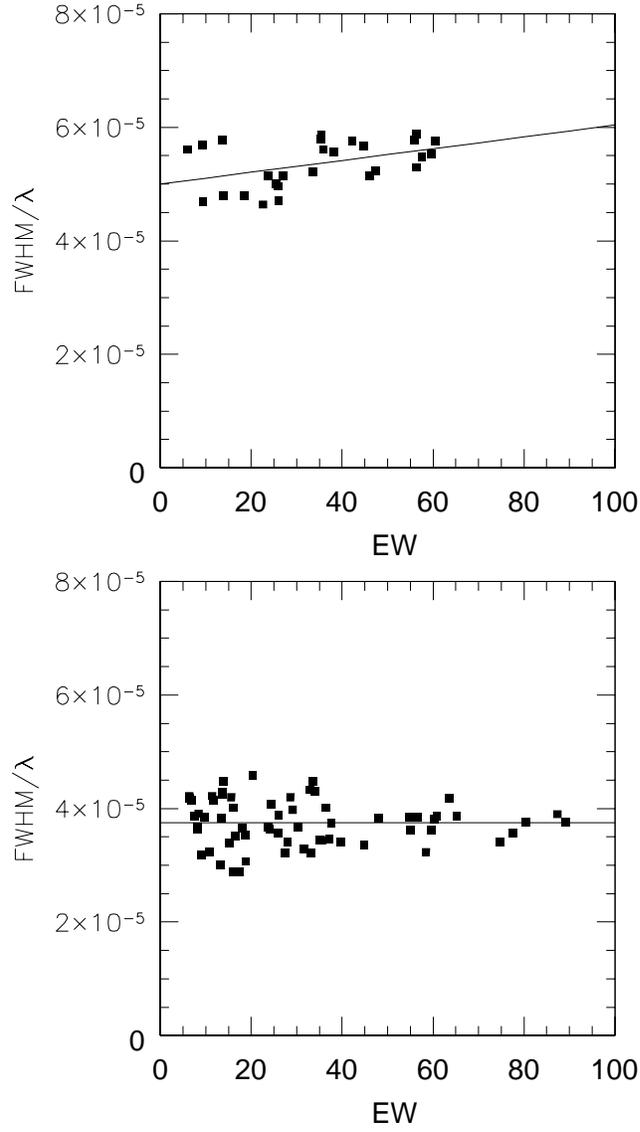}
\caption[ecarretta_fig01.ps]{The fiducial relation between the FWHM of lines and their EWs 
is shown for
stars HE~0024$-$2523 (upper panel) and HE~0130$-$2303 (lower panel).
\label{fig_rel}}
\end{figure}

The lines were then measured by a Gaussian fitting routine using a small
spectral region (of width 1.6 times the FWHM) centered on their expected
location, based on a preliminary determination 
of the geocentric radial velocity.  A
number of lines were discarded at this point based on several criteria 
(features that were not well centered, that were either too broad 
or too narrow, etc.).
After this first measurement, a relation was set between equivalent width and
FWHM for each spectrum, examples of which
are shown in Figure~\ref{fig_rel}.
This relation was then used to obtain a better
determination for each absorption line, 
invoking a different Gaussian fitting
routine.  This second routine has only one free parameter for each line,
effectively the central depth of the line profile, 
because the line center
location is fixed by the average radial velocity determined
from all the lines measured in the second step.  Again, several criteria were
used to discard lines at this point in the analysis (e.g., asymmetric error
distributions, indicating lines that are not well centered; large residuals
compared with expected noise, etc.).  These procedures allow us to obtain very
stable measures of the EWs, with random errors close to those expected from
photon-noise statistics (Cayrel 1989).  Of course, systematic errors may
still be present, in particular those related to the adopted reference continua
and the relation between equivalent width and FWHM.

Table~\ref{t:ewbright} and Table~\ref{t:ewhes} list the final values of the
EWs, along with the adopted atomic parameters for all lines in our list.
Table~\ref{t:nist} in the Appendix summarizes the comparison between our
adopted $gf$'s and the compilation in the NIST database.

In order to evaluate the internal errors in our measurement of EWs, we compare
values obtained for two stars in our sample with similar physical parameters.
We performed this comparison using two dwarfs (HE~0130$-$2303 with \teff,
\grav\ , [Fe/H], $v_t$ = 6560/4.3/$-$2.96/1.39, and HE~0148$-$2611 with \teff,
\grav\ , [Fe/H], $v_t$ = 6550/4.3/$-$3.07/1.25, respectively) and two giants
(CS~22950-046: \teff, \grav\ , [Fe/H], $v_t$ = 4730/1.3/$-$3.30/2.02, and
CS~22878-101: \teff, \grav\ , [Fe/H], $v_t$ = 4775/1.3/$-$3.09/2.01).

\begin{figure}
\epsscale{1.0}
\plotone{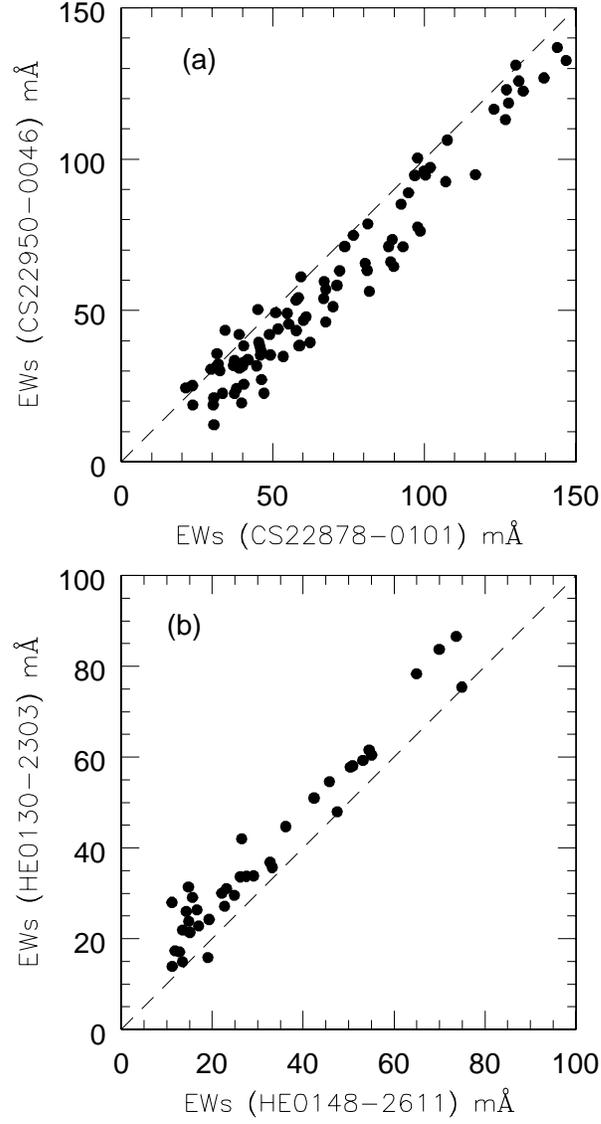}
\caption[ecarretta_fig02.ps]{Comparison for the sets of EWs measured in a pair of giants
(panel a) and in a pair of dwarfs (panel b) of the present study, with similar 
atmospheric parameters.
\label{fig_compint}}
\end{figure}

For the two dwarfs, the r.m.s. scatter about the regression line between the
sets of EWs (see Figure~\ref{fig_compint}, panel (b)) is 4.5\,m{\AA}.  If we
assume that both sets of EWs have equal errors, we can estimate typical errors
of 3.2\,m{\AA} in the EWs. For the giants (Figure~\ref{fig_compint}, panel a),
the r.m.s. scatter is 7.9\,m{\AA}, corresponding to an error of 5.6\,m{\AA} in
the EWs for each star.  In both cases, these errors are in good agreement with
the predicted errors obtained from the formulae derived by Cayrel (1989),
given the spectral resolution and the S/N characteristics of our data.  This
confirms the high quality of the spectra and suggests that no extra sources of
noise were introduced by the EW extraction procedure.

An external comparison of our derived equivalent widths for stars in our
sample can be carried out using the bright, well-studied metal-poor stars,
as well as the stars selected from HK survey.

\begin{figure}
\epsscale{1.0}
\plotone{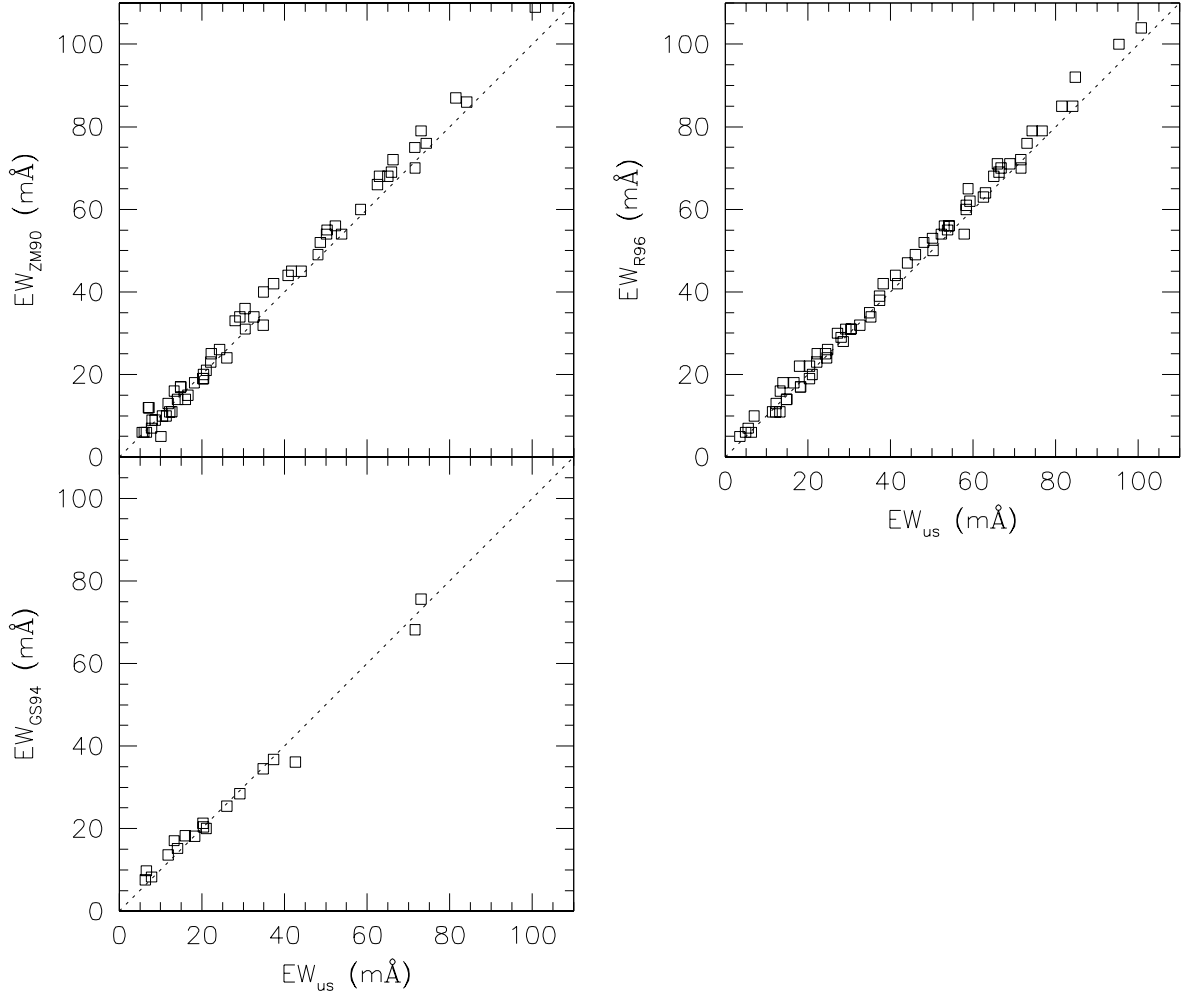}
\caption[ecarretta_fig03.ps]{Comparison for the sets of EWs measured in HD 140283 
in the
present study with the works of Zhao \& Magain (1990; upper panel on the 
left), Gratton
\& Sneden (1994; lower-left panel) and Ryan \etal\ (1996; right panel). The 
dashed line indicates equality.
\label{fig_compext}}
\end{figure}

HD~140283 is the star with the largest number of entries in the Cayrel de
Strobel \etal\ (2001) catalog (note that 
our data for this star are of higher
quality than those for our typical program stars). Among the large list of
previous analyses of this star we considered three sets of high quality EWs:
Zhao \& Magain (1990; $R \sim 20,000$; $S/N \sim 100$), Gratton \& Sneden 
(1994; $R \sim 50,000$; $S/N \geq 150$) and Ryan, Norris, \& Beers
(1996; hereafter RNB96; $R \sim 40,000$; $S/N \sim 45$). The comparison 
is shown in the three panels of
Figure~\ref{fig_compext}, where the one-to-one correspondance lines are also
displayed. The linear regression lines are: EW$_{us} = 0.6$\,m{\AA}+ 0.94($\pm
0.01)$EW$_{ZM90}$ , with $\sigma=2.2$\,{\AA} from 59 lines in common, EW$_{us} =
-1.4$\,m{\AA} + 1.04($\pm 0.01)$EW$_{GS94}$, with $\sigma=2.3$\,m{\AA} from 18 lines in
common, and EW$_{us} = 1.22$\,m{\AA} + 0.93($\pm 0.01)$EW$_{RNB96}$ , with
$\sigma=2.3$\,m{\AA} from 74 lines in common.  The very small scatter present in
these comparisons again agrees well with the theoretically predicted errors,
and confirms the error estimates given above.

For the two giants from the HK Survey in common with the McW95 sample, the
linear regression between our measurements and their EWs is EW$_{McW95} = 1.6$
m\AA\ + 1.04($\pm 0.03)$EW$_{us}$, with $\sigma=12.0$\,m{\AA} from 153 lines in the
two stars. The McW95 EWs are on average larger than ours (the average offset is
4.1 m\AA);  the difference increases for stronger lines. The rather large
scatter is most likely due to their measurement errors, arising from spectra
having a lower resolving power (R$\sim$ 22,000) and a lower average S/N
(typically $30$--$40$) than ours.

\section{ABUNDANCE ANALYSIS}

Model atmospheres with appropriate parameters (see
Table~\ref{t:atm}) were extracted from the grid of Kurucz (1995) with the
overshooting option switched off, interpolating among the nearest models in the
Kurucz grid. The abundance analysis was performed using measured EWs
(Table~\ref{t:ewbright} and Table~\ref{t:ewhes}); the resulting abundances and
element ratios for each species in each star are listed in Table~\ref{t:el1}
and Table~\ref{t:el2}, and are discussed below . In these tables n is the
number of lines used in the analysis of a given ion, and $\sigma$ is the r.m.s.
scatter in abundance from individual lines for the set of lines used for a
particular ion.

The abundances of neutral species are computed with respect to Fe I, while
singly ionized species are compared to Fe II abundances to decrease
uncertainties due to the choice of atmospheric parameters.

Table~\ref{t:sens} shows how the choice of a given set of atmospheric
parameters might affect the derived abundances. Values in this table are
computed by varying, one at a time, the individual atmospheric parameters and
comparing the resulting abundances with the original values. The amount of the
variation is set by the uncertainties established for each of the parameters.
This exercise was carried out for a giant (CS~22950-046) and for a dwarf
(HE~2344$-$2800), in order to span the whole range of \teff\ sampled by program
stars. In each case, the first four columns show the sensitivity of abundance
ratios to changes in each adopted parameter.

The last column of Table~\ref{t:sens}
lists for each star the sums (in quadrature) of contributions
due to the individual parameters; this provides an estimate of the overall
uncertainty in abundance for each species arising from errors in the
adopted atmospheric parameters.  From Paper I we derive estimates of our
internal errors of about $\pm 100$ K and $\pm 0.2-0.3$ dex for \teff\ and for
\grav\ respectively.  In evaluating the total errors, we took into account the
correlation between the error in \teff\ and the error in \grav\ for each star
which arises due to the procedure we adopt to derive surface gravities.

Internal errors in the microturbulent velocity can be checked by using the
errors of the relationship between the abundances of Fe I and the expected line
strengths. Given the rather large number of measured Fe I lines and the wide
range spanned by their intensities, the internal uncertainties in $v_t$ are
quite small ($\pm 0.12$ km/s), so that the 0.2 km/s adopted in
Table~\ref{t:sens} can be viewed as a conservative estimate. Unlike abundance
analyses of solar metallicity stars in this range of \teff, in these very
metal-poor stars the term arising from uncertainties in the microturbulent
velocity ($v_t$) does not dominate the abundance errors due to the general
overall weakness of the lines (save for a few isolated specific cases such as
the Mg b and Sr II lines).

In the remaining part of this Section we discuss some relevant features of our
analysis.

\subsection{Iron}

Iron abundances for our program stars are summarized in Table~\ref{t:el1}.
There were no credible detections of any Fe II lines in the spectrum
of one dwarf (HE~0218$-$2738) from the HES sample. 

As we will see in the next subsection, the scatter in the differences between
abundances derived from Fe~I and by Fe~II lines is quite small; we then feel
justified in assuming a constant offset between the Fe~II and Fe~I abundances
of stars in our sample in order to obtain a value for the [Fe~II/H] ratio for
the one star with no measured Fe~II lines (HE~0218$-$2738).  For this star
only, 
we set $\mbox{Fe~II}-\mbox{Fe~I}=-0.06$\,dex. Abundance ratios of singly
ionized species for this star given in Table~\ref{t:el1} and Table~\ref{t:el2}
are then referred to the abundance of Fe~II obtained in this way.

According to the strict definition given in Paper I, six stars in the sample
can be considered true EMP stars ([Fe/H]$<-3$): the two giants from the HK
survey, and four dwarfs from the HES sample.  Two other stars, BS~17447-029
and HE~0130$-$2303, are borderline, following this definition.

\subsubsection{Uncertainties in Fe Abundances \label{section_unc}}

\begin{figure}
\epsscale{1.0}
\plotone{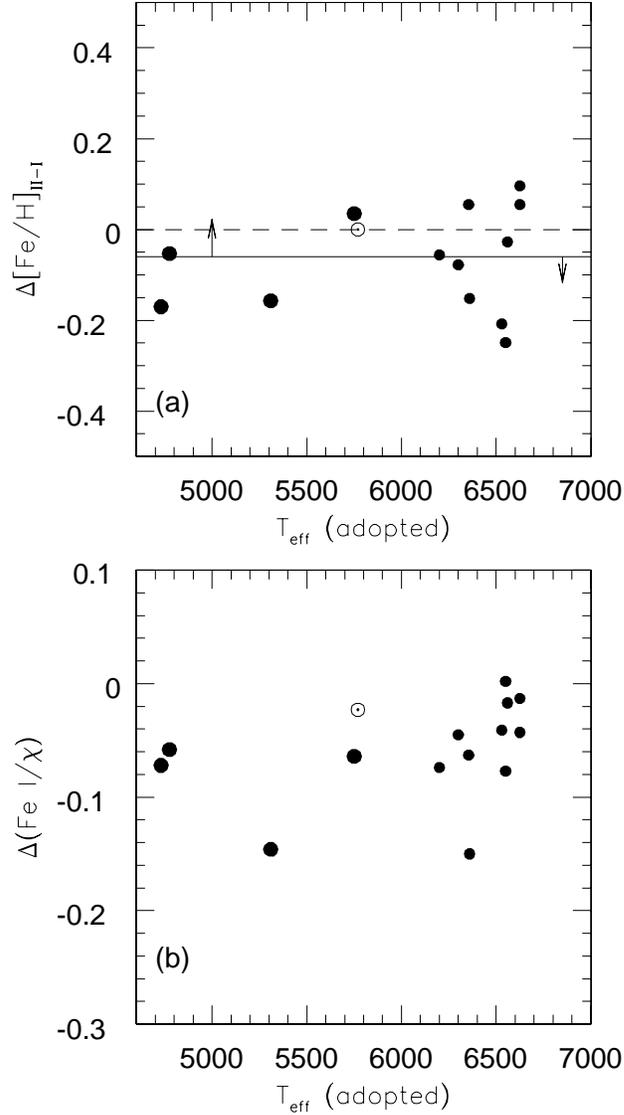}
\caption[ecarretta_fig04.ps]{Panel (a): differences of abundances of iron from singly
ionized and neutral lines as a function of adopted temperatures for program
stars. The dashed line is the value for the Sun, while the solid line is 
the average
value from dwarfs in out sample. Arrows on the left and on the right
side of the Figure
indicate the change in Fe~II$-$Fe~I resulting from a change of +100 K in \teff\,
for a giant and a dwarf respectively.
Panel (b): the slope of the relationships of iron abundances logn(Fe~I)
versus excitation potential (E.P.) as a function of the adopted \teff\ . 
The circled dot is the position of the Sun in these plots.
In both
panels (and in following figures) larger symbols indicate giant and/or subgiant 
stars in our sample, i.e. stars with surface gravity greater than \grav\ = 3.8
\label{fig_dtheta}}
\end{figure}

Among the various diagnostics that can be used to test our Fe abundances, we
considered the differences between the abundances derived from neutral and
singly ionized Fe lines (hereinafter \diff), and the slopes of the
abundances derived from neutral Fe lines with excitation potential (hereafter,
$\Delta(Fe~I/\chi)$). (See Figure~\ref{fig_dtheta} a,b, where both these quantities,
derived for each star, are plotted against effective temperature.  Values of
these parameters are listed for each stars in the last two columns 
of Table~\ref{t:atm}.) These
diagnostics are useful because both temperatures and gravities were derived
independently of our line data (note however that we adjusted $v_t$ in order to
reproduce similar abundances from weak and strong lines). Ideally, we
expect both \diff\ and $\Delta(Fe~I/\chi)$ to be null. However, there are various
reasons why this might not occur in practice: (i) the atomic parameters
adopted in our line analysis may themselves contain systematic offsets or
trends; (ii) our adopted effective temperature scale, or the 
theoretical isochrone
used to derive gravities, may be incorrect; (iii) the 1-d theoretical
constant-flux model atmospheres used throughout this paper may be not an
adequate representation of the real stellar atmospheres; (iv) departures from
LTE in the formation of Fe lines may significantly affect the derived
abundances; and (v) observational errors both in the colors (affecting
individual temperatures) and in the equivalent widths may introduce significant
scatter. We leave aside here other possibilities, such as binarity of
some stars, that might be used to explain individual discrepant points.

Clearly the list of possible concerns is long. It is thus not surprising that
both \diff\ and $\Delta(Fe~I/\chi)$\ exhibit definite zero-point offsets, and
also possible trends with effective temperature (or, equivalently, luminosity).
On average we have $<$\diff$>=-0.07$ dex with $\sigma$(\diff) = 0.11 dex (in
the sense that abundances from neutral Fe lines are larger than those obtained
from singly ionized Fe lines; the error value is the r.m.s. scatter of the
individual values for each star), and $<\Delta(Fe~I/\chi)>=-0.046$ dex/eV with 
$\sigma[\Delta(Fe~I/\chi)] =
0.025$~dex/eV (where we have excluded HE~0132$-$2429 and HE~0242$-$0732, two
stars that show obvious large trends of abundances with line excitation, and
which will be discussed later). If we exclude the two giants,
the average value of \diff\
becomes $-0.06 \pm 0.03$ dex\footnote{The difference is
slightly larger for the two giants, being -0.11 dex.}, with
a $\sigma=0.11$ dex (11 stars).  To quantify these values in terms of 
uncertainties in the atmospheric parameters, as suggested by the referee, we 
note that a difference in $\Delta(Fe~I/\chi)$ of
0.05 dex/eV corresponds to a temperature difference of $\sim250$ K, and that
a difference of 0.07 dex in the iron abundances derived from Fe I and Fe II 
implies a difference of about 0.16 dex in \grav.

When considering the implication of these tests, several points should be taken
into account:

\begin{itemize}

\item The Fe oscillator strengths we used throughout this paper are the best
determinations we found in the literature; they are discussed at length in a
number of papers devoted to the solar Fe abundance (see e.g. Asplund \etal\
2000, and references therein). The transition probabilities for the
Fe~I lines were used as published,
in spite of the fact that a small offset may exist
between values obtained from the absorption experiments of the Oxford group,
and those from the selected laser induced excitation experiments by the 
Hannover
group. (A line-to-line comparison shows that these latter values are larger on
average by $\sim 0.03$~dex.)  Furthermore, the zero point of Fe~II
oscillator strengths is not yet firmly established, and possible offsets of
several hundredths of a dex are easily possible (see the discussion in Asplund
\etal\ 2000).

\item Even using the best available oscillator strengths, small offsets in
the abundances are present also in a solar analysis done following precepts
similar to those adopted for the program stars (LTE, 1-d model atmospheres from
the Kurucz CD-ROMs, etc.) and $EW$s from Rutten and Van der Zalm (1984). In
this case we find Fe abundances of $\log n(Fe)=7.512\pm 0.012$\ (with an
r.m.s. scatter of 0.069 dex for individual lines) from 34 Fe~I lines, and
$7.450\pm 0.016$\ (r.m.s. scatter of 0.085 dex) from 27 Fe~II lines (only lines
with $EW<100$~m\AA\ were considered, to reduce concerns related to the handling
of collisional damping\footnote{Throughout this paper, collisional damping 
was considered by multiplying the Van der Waals broadening by an enhancement
factor given by logE=log(1+0.67 E.P.)(Simmons \& Blackwell 1982).}). 
In view of the roughness of the methods used, these
abundances compare quite well with the much more sophisticated results obtained
by most recent analyses of the solar photospheric spectrum (e.g., Asplund
\etal\ 2000). Our solar analysis yields 
$<\Delta(Fe~I/\chi)> =-0.023$ dex/ev with
$\sigma(\Delta(Fe~I/\chi)) = 0.009$~dex/eV, which is quite 
similar to the value we
obtain for our very metal-poor program stars. Note, however, that the set of
lines used in this solar analysis is disjoint from that used for the stellar
analyses, because those lines measurable in extremely metal-poor stars are
heavily saturated in the solar spectrum. Line selection is of particular
importance for the determination of $\Delta(Fe~I/\chi)$, which is also 
sensitive to the
value adopted for the microturbulent velocity. On the other hand, trends of
abundances with line excitation have been found also in previous, much more
accurate analyses of the solar photospheric spectrum (see e.g. Grevesse \&
Sauval 1999). In this respect, it is interesting to note that such a trend is
not present when 3-d models are used to analyse Fe lines in the solar spectrum
(Asplund \etal\ 2000). 3-d effects may be expected to have an even larger
impact on the analysis of metal-poor stars than in the case of the Sun 
(see Asplund \etal\ 1999). In
fact, due to the lower opacity in the atmospheres, the lines are expected to
form deeper in the stars, where the impact of convection is larger, and
possibly the  granulation contrast larger as well.

\item The r.m.s. scatter of the individual values of \diff\ (0.11 dex) 
roughly agrees with the
expected uncertainties. To show this, we assumed that the internal errors in
\teff\ are those estimated in Paper I, i.e. $\pm 75$ K, possibly increasing up
to $\pm 150$ K for turnoff stars only. To estimate the corresponding errors in
\grav, one needs the slope of the isochrone in the region of interest,
namely $\Delta$(\grav/\teff) = $\sim$0.003 dex/K along the RGB and much less
($\sim$0.0005 dex/K) for less-evolved evolutionary stages (subgiants and
dwarfs). Below the main-sequence turnoff the slope reverses in sign. Reading
from Table~\ref{t:sens} the changes in Fe~I and Fe~II abundances expected for a
100 K increase in \teff\ (with consequent changes in \grav\ and overall
metallicity [A/H]), the resulting value of \diff\ would be about +0.08 dex
for giants and $-0.04/-0.08$ dex for subgiants/dwarfs. When we sum these values
in quadrature with the internal error due to uncertainties in the measured 
EWs, for our Fe~II abundances (typically
0.05~dex for giants and 0.09~dex  for dwarfs, where only a few lines were
usually observed), we find that the observed scatter in the values of \diff\
is reproduced.

\item An interesting feature of our analysis is that adoption of a
systematically incorrect \teff\ scale would, as described above, 
yield values of \diff\ of opposite
signs for dwarfs and giants when the resulting changes in \grav\
are included. 
However, a
larger sample is probably needed to explore this possibility. Again, adoption
of realistic 3-d model atmospheres for the program stars might provide better
insight into this issue (Asplund \etal\ 1999).
\end{itemize}

\begin{figure}
\epsscale{1.0}
\plotone{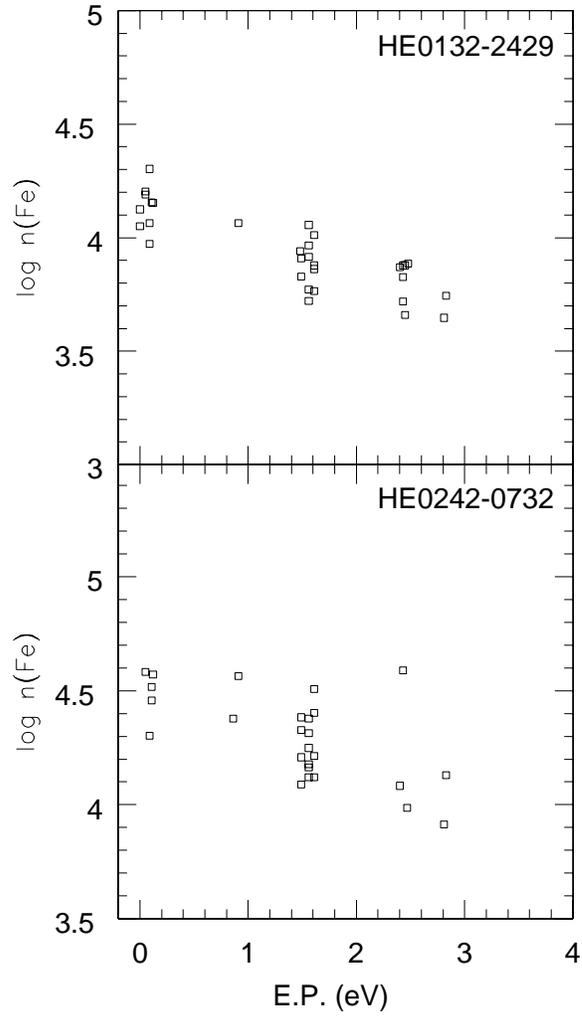}
\caption[ecarretta_fig05.ps]{Run of the iron abundances with excitation potential (E.P.) for
stars HE~0132$-$2429 and HE~0242$-$0732,
the two stars that appear anomalously low
in Figure~\ref{fig_dtheta}.
\label{fig_feep}}
\end{figure}

For most of the sample stars, the general trends shown in
Figure~\ref{fig_dtheta} can probably be explained by \teff\ scale errors or
model atmosphere uncertainties.  However, two stars (HE~0132$-$2429 and
HE~0242$-$0732) display trends of abundance with excitation potential
($\Delta(Fe~I/\chi)\leq -0.15$) that are much larger than can be induced by typical
errors in the stellar parameters, primarily in the \teff.  The trend found for
HE~0132$-$2429 is clearly significant, and does not depend on inclusion or not
of one or two lines (see Figure~\ref{fig_feep}).  The situation for
HE~0242$-$0732 was less clear since most of the 
apparent trend  obtained when the atmospheric parameters of Paper I 
were used for this star 
was due to a single line. 

In order to remove these trends, \teff's much lower (by $\sim 500$~K) than
those given by the colors would be required. Such differences cannot be due to
the reddening corrections, because these are too small.  Furthermore, \teff's
derived from the H$_\delta$\ profiles support the high \teff\ values given by
colors (see Paper I: however, we note that H$_\delta$\ is quite weak for
HE~0132$-$2429, so that precise determination of \teff\ is not easy for this
star). Also, it does not seem possible to invoke contamination by a bluer
companion such as main-sequence turnoff star (this explanation may obviously
only be considered for HE~0132$-$2429). To demonstrate this, we note that the
observed (de-reddened) $V-K$\ color for this star is $V-K=1.86$. This color
can be obtained by combining the flux from a star along the RGB (with
$M_V=1.4$, $V-K=1.91$) and a star at the main-sequence turnoff ($M_V=3.8$,
$V-K=1.05$). In the (observed) blue part of the spectrum, the secondary would
be about 7 times less luminous than the primary: it would be difficult to
detect such a star if the radial velocity difference were small.  However, the
difference between the color of the primary and that of the whole system would
be only 0.05 mag in $V-K$, corresponding to less than 100 K in \teff. This
difference is far too small to explain the trend of Figure~\ref{fig_feep}.

In the case of HE~0242$-$0732, the automatic EW measuring routine did not
pick up any Fe~II lines.  Hence we measured the EWs of the four strongest Fe
II lines expected to be present in our spectra, at 4233.16, 4923.93, 5018.45
and 5169.03\,{\AA}, by hand.  All these lines are only a bit stronger than
noise, so that their EWs are somewhat uncertain; however, they surely are not
much stronger than our estimates.  Using these lines, a lower limit for the
surface gravity may be obtained by assuming that log n(Fe~I) $\sim$ log
n(Fe~II).However, the Fe~II lines 
in this star are clearly much weaker than expected on the basis of the
atmospheric parameters used thus far (\teff=6455, \grav=4.2)
from Paper I. This forced us to
consider both a lower temperature, and a less evolved evolutionary phase for
this star. A lower temperature is also suggested by the H$_\delta$ profile.

In Paper I we already noticed that the $K$ magnitude for this star from the 
2MASS is quite uncertain (it is the faintest star in the infrared  in our sample). 
On the whole, it might be reasonable to disregard for this star \teff\ from 
$V-K$, and use only the temperature from 
the more accurate $V-J$ color, that is to adopt a \teff\ of 
6360 K. Also, it may be reasonable to consider a gravity consistent with a MS 
star at this \teff\ (\grav=4.4), rather than a star brighter than the turn-off.

We then repeated the analysis with the revised stellar parameters, after
adding manual measurements of four more weak Fe~I lines, to extend the range
of excitation potential sampled. With the same procedure adopted above, we
derive the full set of parameters for HE~0242$-$0732 of 6360/4.4/$-$3.21/0.40.
The abundance from Fe~II lines is still lower than that obtained from Fe~I
lines, although the difference is roughly halved with these parameters with
respect to the original discrepancy). We regard this new analysis as quite
robust, and we adopt these new parameters for this star.

Using these new parameters and the slightly more extended line list, the
discrepancy of star HE~0242$-$0732 in Figure~\ref{fig_dtheta} is reduced.
There is still a trend
of abundances with E.P., but it is well within the error bars, which are
large, due to the small range in E.P. of available Fe~I lines. The trend is
not very different from the trends shown by the other stars in the sample.
Another improvement achieved using this set of atmospheric parameters is that
the Ti II abundance (that was previously the lowest in the sample) now is
raised to +0.25 dex, and does not stand out anymore as outlier in
Figure~\ref{fig_alpha2} (see below).

\subsubsection{Comparison with results from a purely spectroscopic analysis}

As a further check of our adopted procedure to derive atmospheric parameters,
we performed a purely spectroscopic analysis, deriving temperatures, surface
gravities, overall metal abundance and microturbulent velocities from the
equivalent widths alone.  In this process, we ignored the photometry
that forms the basis of the assignment of stellar parameters used
in Paper~I.

We adopted HE~0242$-$0732 as typical of our program stars (it is neither the
best nor the worst case, as far as the quality of observational material is
concerned); however it does represent an extreme case of differences between
results of our original analysis (with temperatures based on colours, and
gravities from the color-magnitude diagram), and those from a purely
spectroscopic analysis. The spectroscopic set of parameters in this case is:
an effective temperature \teff=5250\,K (obtained by zeroing the slope in the
relation of Fe abundances with the excitation potential of lines), a surface
gravity \grav = 2.55 (assuming that Fe I and Fe II lines must give the same
iron abundance), [A/H]$=-4.21$ (given by the abundance of iron) and a
microturbulent velocity of 0.45 km/sec (derived eliminating trends of
abundances with expected line strength for Fe I lines).

This set of parameters is quite different from that of our original analysis.
To estimate the appropriate error bars, we must take into account the 
covariance among the errors, and hence we performed the following exercise.
Using this set as a starting point, we used the same set of Fe lines to repeat
the analysis changing the $v_t$\ value until the 1$\sigma$ value from the slope
of the abundance/line strength relation was reached, and optimizing at the same
time the other parameters at their best value. The resulting set (\teff,
\grav, [A/H], $v_t$), was: 5480/2.85/-4.07/0.74. Starting then from this second
set of parameters, we changed the effective temperature in order to have all
parameters optimized when the 1$\sigma$ value from the starting slope of
abundances vs E.P. relation was reached (still leaving the trend of abundances
with expected line strength off by 1~$\sigma$\ from its best value). In this
way we get a third set of parameters: 5760/3.45/-3.81/0.78. Finally, we
repeated the same exercise, but now achieving a change of 1$\sigma$ in the
difference between Fe I and Fe II (again, leaving trends of abundances with
expected line strength and excitation off from their best value by 1~$\sigma$):
the resulting final set of parameters was 5780/3.05/-3.79/0.88.

We emphasize here that this last set of parameters is still statistically
acceptable: each of the residual trends is not larger than 
its 1$\sigma$ r.m.s. error.

A simple comparison between this last set and our starting point allows us to
give an estimate of 1$\sigma$ internal errors associated with a purely
spectroscopic analysis; we consider these values more appropriate than those
derived from considering each individual parameter as independent from the
other. Uncertainties are: $\pm 530$~K in \teff, $\pm 0.43$ km/s in $v_t$ and
$\pm 1.3$ dex in \grav. On the whole, the uncertainty in the overall metal
abundance from the purely spectroscopic analysis is $\pm 0.42$ dex! This very
large error bar is more than twice that deduced by assuming that the
atmospheric parameters are independent each other. This is mainly due to the
correlation existing between excitation potential and line strength for Fe~I
lines. We conclude that the intrinsic uncertainties in a purely spectroscopic
approach are too large to secure a robust result when dealing with extremely
metal poor stars, where the number of reliable lines measured is rather small,
particularly for Fe~II,
uncertainties in the EWs of individual lines are not negligible, and the
range in excitation potential is limited.

For these reasons, we think that the procedure we have adopted that fixes
temperature and gravity by independent means works better in our case. The
previous exercise, for example, shows that using only information provided by
Fe lines (as it is done in the purely spectroscopic analysis), we could not
say if HE~0242$-$0732 is an RGB or HB star (5250/2.55), or a subgiant
(5780/3.05).  Note that both these combinations of {\teff} and {\grav} are
compatible with the same isochrone, and would then appear as plausible
solutions.

\subsubsection{Checks on Departures from LTE for Fe}

Having derived gravities from \teff\ and isochrones only, we can in principle
use the ionization balance of Fe in order to check if Fe abundances are
affected by departures from Local Thermodynamic Equilibrium (LTE). This issue 
is the subject of considerable
debate, since Thevenin \& Idiart (1999) suggested that LTE analyses tend to
underestimate the Fe I abundances (by a few tenths of a dex in extremely
metal-poor stars), while Gratton \etal\ (1999) pointed out that large ($>0.2$
dex) departures from LTE are unlikely, and LTE should instead be a very good
assumption for Fe line formation in metal-poor dwarfs.  In view of the large
uncertainties still present in the collisional cross sections (e.g., Gehren
\etal\ 2000), and given the limitations in the adopted model atoms, there is no
way at present to decide from first principles which of these 
analyses is valid.

There is no evidence in our data for a large Fe overionization, judging from
the upper panel in Figure~\ref{fig_dtheta}. As discussed above, the average
difference in abundances from Fe I and Fe II lines is only marginally
different from zero. Moreover, almost all stars show differences which have
the wrong sign for overionization.  This seems to rule out large departures
from LTE in the formation of Fe lines in metal-poor stars. 

On the other hand, the discussion of \S\ref{section_unc} showed that there
are so many possible uncertainties still remaining in the temperature scale,
in the model atmospheres, and in the oscillator strengths, that small NLTE
effects (at a level of 0.1 to 0.2~dex) cannot be firmly excluded from our data.

\subsection{The $\alpha-$elements}

We detected lines arising from two stages of ionization for titanium, hence
we can use the ionization equilibrium of this element as an additional clue for
the presence of NLTE effects.  In fact, given the lower ionization potential
of Ti I, it may be expected that this element ought to be even more vulnerable
to non-LTE effects than is iron.
There is no conclusive evidence for departure
from LTE from the Ti~I/Ti~II abundance. From the data in Table~\ref{t:el2}, a
trend for decreasing Ti~II $-$ Ti~I differences as \teff\ increases seems to be
present, with the two giants showing somewhat larger (positive, i.e.,
abundances from Ti~II larger than those from neutral Ti) differences. However,
even in this case we believe that the safest approach is not to draw any firm
conclusions until a larger sample becomes available. We are concerned that
there is a possible bias resulting from the rather uncertain Ti~I abundances in
the warm, metal-poor dwarfs that are clustered around Ti~II$-$Ti~I $\sim -0.2$ to
$-0.3$ dex.  Moreover, we are not aware of any computations of NLTE effects for
Titanium.

We note in passing that data from McW95, and the few stars from
RNB96 with both Ti~I and Ti~II measured (all giant/subgiant stars) show the
same pattern of differences in titanium abundances as a function of
metallicity, with an average value of Ti~II$-$Ti~I just below zero; values as
low as about $-0.3$ dex can be found.

\subsection{Aluminum}

For Al, the only accessible feature in our HIRES spectra is the resonance
doublet at 3944-3961\,\AA.  However, a reliable measurement of these lines is
somewhat hampered by the proximity of H$_\epsilon$ to the 3961\,{\AA} line.
Moreover, the line at 3944\,\AA\ is disturbed by CH features in some of our
stars (Arpigny \& Magain 1983).

Apart from these problems, this doublet is not an ideal abundance indicator in
view of the presence of possible large departures from LTE, extensively
discussed by Baum\"uller \& Gehren (1997). They found that LTE analyses produce
a large underestimate of Al abundances; they give NLTE corrections (about
0.6 dex) for the resonance lines in their Table~\ref{t:atm}.

Very recently, Gratton \etal\ (2001) found that inclusion of these corrections
improves the agreement between Al abundances derived in globular cluster dwarfs
from this doublet and from the IR high excitation doublet at 8772.9$-$8773.9\,{\AA},
which is believed to be less affected by departures from LTE. This provides
further support for the calculations of Baum\"uller \& Gehren. We therefore
interpolated Table 1 of Baum\"uller \& Gehren to derive corrections to the Al
abundances listed in Table~\ref{t:el1}.

\subsection{The Fe-peak Elements}

Apart from Fe, our spectra provide useful information on the abundances of
three additional Fe-peak elements (Sc, Cr, and Mn). Lines of the remaining
elements are too weak to be reliably measured, due to the combination of low
metal abundance and high temperature of most of our stars.

Lines of Sc and Mn exhibit hyperfine structure (hereafter, HFS), mainly due to
their non-zero nuclear magnetic moments. HFS is expected to be rather narrow
for the Sc II lines, with small impact on abundances, hence we neglect it.
On the other hand, Mn lines exhibit broad HFS: this was taken into account
using data from Booth \etal\ (1983).

\subsection{The Neutron-Capture Elements}

We measured lines for two n-capture elements (Sr and Ba; in both cases we
observed the resonance lines of the singly ionized species). The Eu II line at
4129.7\,{\AA} is within the observed spectral region, but it is too weak to be
detectable in any of our spectra. \footnote{For a EMP dwarf near the main
  sequence turnoff with [Fe/H] = $-3.0$, an upper limit of 10\,m{\AA} for the
  EW of the strongest Eu~II line corresponds to [Eu/Fe] $\lesssim +1.8$ dex,
  too large to provide any interesting constraints.}  Note that none of our
stars belongs to the group of rare stars that exhibit strong overabundances of
elements produced by the r-process, which includes CS~22892-052 and
CS~31082-001.  The classification of BS~17447$-$029 is somewhat uncertain;
this star shows high abundances of Sr and Ba, but does not have detected Eu
(nor any strong CH features). This star may be a mildly r-process-enhanced
star.

HFS due to both non-zero magnetic moment and isotope splitting is
significant for the Ba~II line at 4554.04\,{\AA}.
The HFS corrections were evaluated using the data of Steffen (1985); those
recently calculated by McWilliam (1998) are very similar.
Note, however, that we do not know the relative abundances of the
different Ba isotopes: our assumed distribution is the solar one, and may be
inappropriate for metal-poor stars. 

\section{DISCUSSION}

\subsection{Results from the Keck Pilot Program}

\subsubsection{Comparison for Stars in Common with McWilliam \etal\ (1995)}

There are two stars in common between our sample and that analyzed by
McWilliam \etal\ (1995): the two giants CS~22878-101 and CS~22950-046.
The adopted \teff\ are quite similar (ours being larger on average 
by 38 K); we adopted larger surface gravities (on average
by 0.3 dex), and smaller microturbulent velocities (on average by
0.52~km/s).
The effects of these differing choices of \grav\ and of $v_t$ are in
most cases similar in magnitude but
of opposite sign, and so somewhat cancel out.
The differences between our and the McW95 analysis are as follows (in the sense
ours-McW95; in parenthesis we give the difference expected from the atmospheric
parameters): Fe +0.09 (0.13); Mg +0.15 ($-$0.09); Al $-$0.17 (0.11); Si +0.04
($-$0.07); Ca $-0.15 (-0.09$), Sc +0.17 (0.07); Ti I $-0.24 (-0.09$); Ti II
$-$0.04 (0.06); Cr +0.02 (0.01); Mn $-0.19 (0.14)$; Sr +0.20 (0.24); Ba $-0.30
(+0.02)$.

In several cases there is a reasonable agreement between the observed and
expected offsets. The large difference for Mg is due to the $gf$'s we adopted; see
the Appendix.  The differences for Al and Si are discussed elsewhere in the
text. Part of the difference for Sc is due to the fact that we neglected the
HFS for this element (while it is considered by McW95).

\subsubsection{$\alpha-$elements}

\begin{figure}
\epsscale{1.0}
\plotone{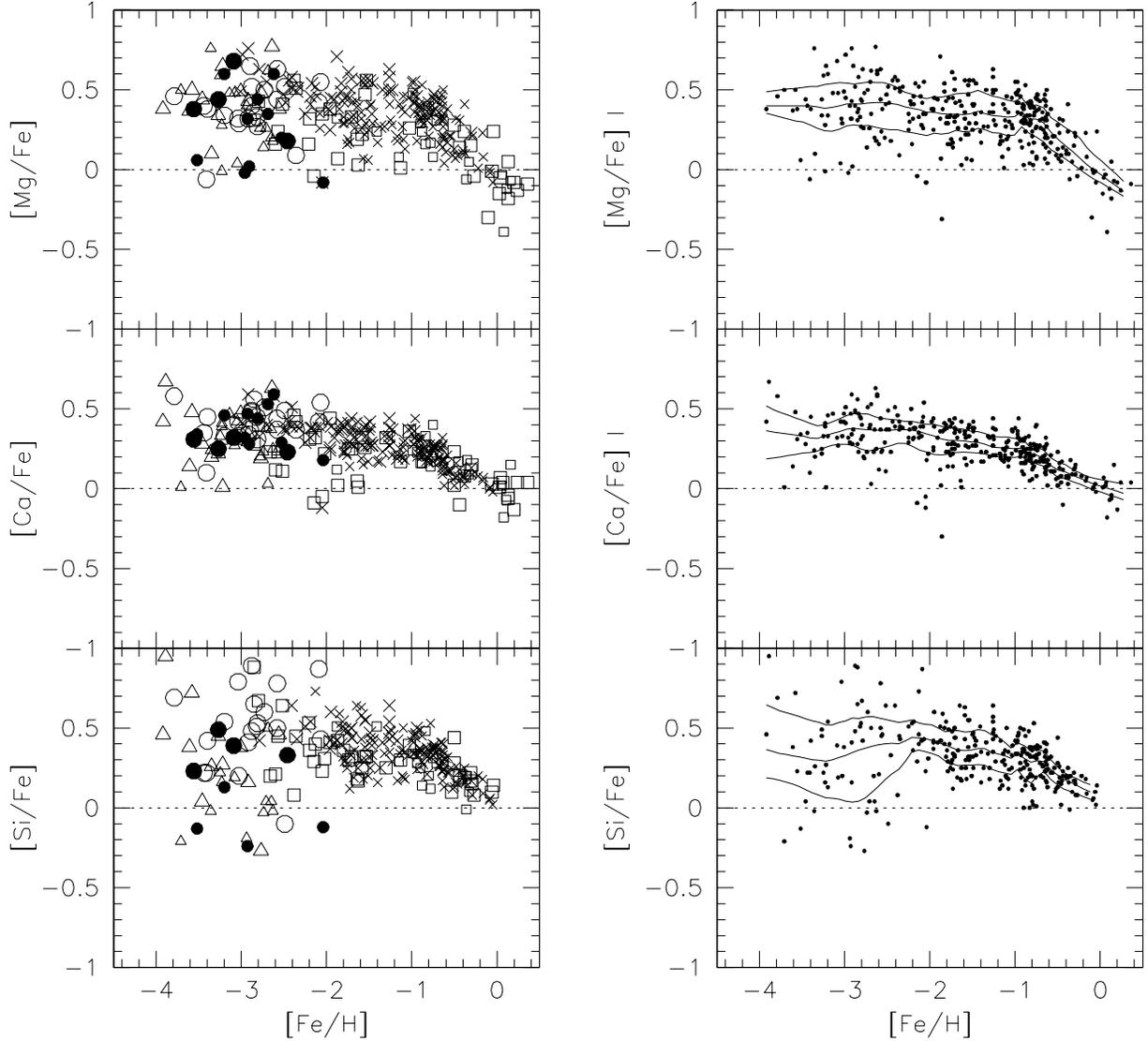}
\caption[ecarretta_fig06.ps]{Run of the $\alpha-$elements Mg, Ca, Si abundances as a 
function of metallicity for program stars (filled circles) as compared to 
other studies (left panels): open circles: McWilliam \etal\ stars;
open triangles: Ryan, Norris, Beers (references in Norris \etal\ 2001); open
squares: Gratton samples (see Norris \etal\ 2001 for references); 
crosses: Stephens (1999). Larger symbols indicate giants stars, smaller symbols
dwarf stars. Right panels show 
data from all the sources listed above, 
additional data from Gilroy \etal\ (1988), Carney \etal\ (1997), Nissen \&
Schuster (1997) and Stephens (1999),
all displayed as small solid points. The superimposed 
solid lines are the {\it loess} lines
described in Section 5.1
\label{fig_alpha1}}
\end{figure}

Abundances for the $\alpha-$elements are summarized in Table~\ref{t:el1} and
Table~\ref{t:el2}, and shown in the left-hand panels of Figure~\ref{fig_alpha1}
and Figure~\ref{fig_alpha2}.  We also show, for purpose of comparison, the
results from the literature of
previous high-dispersion studies: McWilliam \etal\
(1995) for a sample of 33 metal-poor giants, and a compilation of literature
data taken from NRB2001. In addition to the five EMP giants they studied,
NRB2001 used high-quality data compiled from the literature to discuss the
chemical evolution of the Galaxy.  In the construction of Figures 5--9, 11, 12
and of Table~\ref{t:scat} we employ three datasets from that paper: (i) the
data from the Ryan/Norris/Beers group (collectively RNB hereafter): Ryan,
Norris \& Bessell (1991); Ryan, Norris \& Beers (1996);
Norris, Beers \& Ryan (2000); NRB2001; (ii) the data from Gratton/Sneden
(hereafter GS): Gratton (1989); Gratton \& Sneden (1987, 1988, 1991, 1994); and
(iii) Stephens (1999).

Following Table~\ref{t:atm} in NRB2001, data from McW95 come from lower quality
spectra, but $\alpha-$elements have strong lines in metal-poor stars, hence
concerns due to somewhat lower resolution and S/N are limited in this case. The
overall patterns for each of the $\alpha-$elements are quite similar, even
though our sample is mostly composed of dwarfs. Unweighted mean values for
[Mg/Fe], [Ca/Fe], [Ti/Fe]I and [Ti/Fe]II are respectively 0.31
($\sigma=0.24$)\footnote{Throughout this paper, the symbol $\sigma$ will
indicate the standard deviation of a single measurement, while the value after
the symbol $\pm$ will refer to the standard deviation of the mean.}, 0.35
($\sigma=0.12$), 0.40 ($\sigma=0.19$), 0.36 ($\sigma=0.15$) dex, to be compared
with 0.40 ($\sigma=0.22$), 0.44 ($\sigma=0.13$), 0.32 ($\sigma=0.19$) and 0.30
($\sigma=0.13$) in the McW95 sample.

The average difference in the sense [$\alpha$/Fe]$_{us}$ $-$
[$\alpha$/Fe]$_{McW95}$ is only $-0.01, \sigma=0.08$ dex, suggesting that there
is no significant offset in the mean $\alpha-$element abundances between
metal-poor dwarfs and giants.

The only apparent offset seems to lie in the Si abundances. However, our
[Si/Fe] values are based on only one strong line, Si I 3905\,{\AA}, and this
line was disregarded by McW95 since it exceeded their adopted reduced
equivalent width limit. Unfortunately, all other Si lines in the spectral
region sampled by our spectra are vanishingly weak in these warm metal-poor
stars.  The large scatter in the plot [Si/Fe] vs. [Fe/H] is present in all
samples of metal-poor stars, but the observational problems suggest that this
could be an artifact of the still uncertain abundance analysis for this ion.

We note that 4 out of 14 stars in our sample exhibit distinctly low [Mg/Fe]
ratios, close to the solar ratio.  It would be interesting to have more insight
into these objects by coupling chemical data with kinematical information. This
is deemed important, since more and more stars which show underabundances among
the $\alpha$-elements are being discovered by several investigators, e.g.,
stars with [Fe/H]$<-2$ and [Mg/Fe]$<0.1$ in the upper-left panel of
Figure~\ref{fig_alpha1}. All our low-Mg stars are dwarfs, but they are equally
intermingled with giants in other samples. Often, when orbital parameters can
be estimated, these objects are seen to have large apogalacticon distances,
suggesting their possible origin in lower mass stellar systems with chemical
evolution histories that were distinctly different from that of our Galaxy, and
were subsequently accreted into our Galaxy (see King 1997; Nissen \& 
Schuster 1997; Carney \etal\ 1997).

At present we do not have good information about Galactic orbits for all the
objects in our sample, but the expected extension of the USNO CCD Astrograph
Catalog (Zacharias \etal\ 2000) from the southern sky to the northern sky, as
well as the forthcoming release of proper motions from the GSC-II catalog (and
other ongoing proper motion programs) should fill the 6-dimensional parameter
space, when coupled to the radial velocity information. We thus defer the study
of possible correlations of chemistry and kinematics to future works.

The scatter seen in Ca abundances is similar to the one in McW95 stars, if one
disregards the star with [Ca/Fe]$\sim 0.9$ in their sample, and seems to be
lower than the scatter observed in the RNB data. Our element ratios [Ca/Fe]
seem to be slightly lower, on average, than those from McW95, and more in
agreement with data from RNB, even if we do not find stars having very low Ca
abundances ([Ca/Fe]$<0.1$). Excluding the four most metal-poor stars in 
the McW95 sample, the other datasets do not show any trend for Ca abundances
increasing for decreasing metal abundances.

\subsubsection{Aluminum}

The NLTE Al abundances ([Al/Fe]$_{NLTE}$) for the program stars are shown
in Figure~\ref{fig_al} (central panel) as a function of [Fe/H]. Al abundances
derived without these corrections are shown in the upper panel of this Figure.
In order to make a meaningful comparison, the NLTE corrections from Baum\"uller
\& Gehren (1997) were applied as well to the abundances in the samples from
the literature, also shown in Figure~\ref{fig_al}, with the same symbols as in
previous Figures.

Our average [Al/Fe] value without correction for NLTE is $-0.59 (\sigma=0.15)$
dex, which is not too far from the mean value as defined at [Fe/H]$\sim -3$ by
the robust trend computed in RNB96 (their Figure 3c). This means that as far
the EWs of these resonance lines are concerned, the agreement between the two
different investigations is quite good.  Moreover, an offset in the two
metallicity scales is not likely to be present, even if we have only one star
in common, the well studied HD 140283.  We considered for this star the
differences in atmospheric parameters \teff\, \grav\ and $v_t$ (5750/3.40/1.40,
RNB96; and 5750/3.67/1.10, this study). Reading from Table~3 of RNB96 the
changes in [Fe/H] associated with these differences, the metallicity given by
RNB96 ([Fe/H]$=-2.54$) would increase by only 0.03 dex. We can say that part of
the difference with our value [Fe/H]$=-2.44$ can be explained by differences in
the EWs and part ($\sim 0.03$ dex) by a different distribution 
over excitation potential of 
the sample of lines used in the analysis.

To compare our results with models of chemical evolution, we adopt, based on
theoretical and observational reasons discussed above, the abundances for Al
derived with the NLTE corrections included.  In this case, our average values
[Al/Fe]$= +0.10 (\sigma=0.22)$ and [Al/Mg]$= -0.21 (\sigma=0.23)$ are not
dramatically different from the prediction of models by 
Timmes \etal\ (1995).

The case for the sample of McW95 looks quite different. A large scatter in Al
abundances is seen among their stars at each metallicity, and the reason is not
obvious. We note that all stars in that sample are giants, and the
derivation of atmospheric parameters is somewhat more uncertain for giants than
for dwarfs.

In conclusion, we have to agree with RNB96 that the higher average [Al/Fe]
ratio found by McW95 has no obvious explanation.

\subsubsection{The Iron-Peak Elements: Sc, Cr, Mn}

Our results for the iron peak elements are shown in the left three panels of
Figure~\ref{fig_ironpeak}. In the region $-2<$[Fe/H]$<-1$, the Cr and Mn
abundances show a relatively small scatter as is expected when the
contributions from many SN events average over the IMF and yields (see
Figure~\ref{fig_ironpeak}).
On the other hand, below [Fe/H]$\sim -2$ the scatter increases, which is
characteristic of stochastic models of chemical enrichment. The correlation
in the upper panel (Cr increasing as Mn increases) could be explained either
if the production of both elements is a function of the mass cut (since they
have slightly different atomic number, they are produced in nearby, yet
different regions) or if there is a different neutron excess in this
metallicity range.

A deeper insight can be obtained looking at Figure~\ref{fig_crmn}, where we
plot in the upper panel the ratios [Cr/Fe] vs. the abundance ratios [Mn/Fe] for
stars in the collected ``big sample'', described in \S\ref{bigsample}, with
[Fe/H]$<-2$.  In this low-metallicity regime we expect to see the classical
signature of Type II SN nucleosynthesis. 
However, since the neutron excess is known to be a function of metallicity, we
can test this hypothesis by removing the trend with metal abundance.  To
this end, we used two linear regressions to fit the [Cr/Fe] and [Mn/Fe] vs.
[Fe/H] distributions (for stars with [Fe/H]$<-2$) and computed the residuals of
the abundance ratios with respect to these two fits.  The residuals are shown
in the lower panel of Figure~\ref{fig_crmn}. 

Again, a correlation is evident for both our program stars and for the stars
in other samples from literature. 
Note that the evidence
for a correlation holds even if we disregard the extreme case of
HE~2344$-$2800, with its large Mn~I abundance.  This star has a high-S/N HIRES
spectrum and the high [Mn/Fe] ratio in this star is real beyond any doubt: in
Figure~\ref{fig_permn} we compare the spectrum of this star in the region
around the Mn lines at 4030-34\,{\AA} with the spectrum of HE~0024$-$2523,
with very similar atmospheric parameters, but no measurable Mn lines.

What do these findings tell us? Following Heger \& Woosley (2001) and Qian \&
Wasserburg (2001a,b,c), very massive stars (VMS) that belong to the Population
III explode as pair-instability supernovae and are expected to produce a
[Cr/Fe] ratio approximately constant, while [Mn/Fe] should decrease with
metallicity, since nuclei having odd nuclear charge are underproduced by VMS.
Therefore, the signature of VMS in the iron-peak elements Cr and Mn should be
a lack of correlation.  Since we do observe such a correlation, our results
point toward a likely scenario discussed e.g. by Nakamura \etal\ (1999) who
explain the trends of [Cr/Fe] and [Mn/Fe] ratios decreasing with decreasing
[Fe/H] as due to a variation of mass cuts in type II SN as a function of the
progenitor mass. The trends we observe can be reproduced if the mass cut was
smaller for the larger mass progenitor that presumably was the first to evolve
and pollute the gas in the early halo.

\subsubsection{Heavy Elements: The n-capture Elements Sr and Ba} 

Strontium abundances were derived from the Sr II resonance lines at 4077\,{\AA}\
and 4215\,{\AA}, the other accessible lines being vanishingly weak in 
EMP stars. These strong resonance lines are relatively unaffected by
blends in very metal-poor stars and can be easily measured for most of our
program stars, excepting those with the lowest S/N values.
On the other hand, since these are resonance lines, they are saturated in most
stars. Hence, the derived [Sr/Fe] ratios are sensitive to the details
of the stellar parameters and of the adopted model atmosphere 
(see Table~\ref{t:sens}).  In particular, they
are affected by uncertainties in the microturbulent velocity. However, since we
derive $v_t$ values by using a fair number of Fe I lines in each star, this is
not a serious issue.

Figure~\ref{fig_heavy} presents our results as compared to data from the
literature. In this Figure we plot in the upper panels the run of [Sr/Fe] as a
function of the metallicity [Fe/H] for our program stars and a compilation of
previous studies: McWilliam \etal\ (1995) and RNB for very metal-poor dwarfs
and giants and Gratton \& Sneden (1994; GS94) for somewhat more metal-rich
objects.
While systematic offsets may be present between different studies (mainly
depending on the \teff\ scale adopted and on the set of model atmospheres), the
overall agreement is fairly good.

Our data confirm once more that at metallicities [Fe/H]$\sim -3$ (or below
$-2.4$, the point where most heavy element patterns show a change in slope,
McW95) there is a huge spread in the observed [Sr/Fe] values, reaching almost 3
dex. This scatter is not linked to a particular evolutionary stage, being
present among both dwarf and giant stars.  The increase in scatter for [Sr/Fe]
values for decreasing metallicity is commonly explained by the classical
enrichment of r-process elements from explosions of massive Type II SN (above
12--15\,M$_\odot$) in the framework of a stochastic enrichment mechanism (McW95).
In this scenario, if only few supernovae contribute to the production of Sr, we
expect a strongly asymmetric distribution in the logarithmic plane [Sr/Fe] vs.
[Fe/H], with more stars with low [Sr/Fe] than stars with high Sr.

The r-enhanced region ([Sr/Fe]$>0$ in the upper panels of
Figure~\ref{fig_heavy}) is attributed to the same classical site of r-process
production, but restricted in this case to very few SN less massive than
12--15\,M$_\odot$. As a consequence of the limited number of objects
contributing to the enrichment, the whole process has a strongly stochastic
behaviour revealed in a few stars with abnormally high r-process abundances.

Ba abundances were derived from the Ba II resonance line at 4554.0\,{\AA}. This
line has appreciable HFS, taken into account when computing the abundances of
Table~\ref{t:el2} using data from Steffen (1985).
Figure~\ref{fig_heavy} (lower panels) summarizes and compares our results with
previous studies.  Apart from the few objects above [Ba/Fe]$\sim 1$ (whose
origin is somewhat different, see below), Ba shows a decrease for [Fe/H]$<-2$
and less scatter than the lighter n-capture element Sr.
The main features of this Figure are commonly explained by the two classical
sources of Ba production: the main s-process in intermediate-mass stars
(4--7\,M$_\odot$) evolving through the asymptotic giant branch (AGB) phase, plus
a contribution by the r-process in SN. As a consequence of the evolutionary
stellar timescales involved, what is seen below [Fe/H]$=-2$ is essentially the
enrichment of r-elements from massive Type II SN, whose entire evolution from
birth to death is much quicker than the timescale required for lower-mass
stars to reach the AGB and for complete mixing of the winds from these stars.
 
Finally, Figure~\ref{fig_srba} shows the relation between the abundances of Sr
and Ba -- these elements are often chosen to represent the behavior of light
and heavy n-capture elements, respectively. This plot is quite instructive,
since we can see that, for the majority of the stars, the [Sr/Fe] and [Ba/Fe]
ratios lie almost exactly on a line that is simply that of a scaled solar
composition.  The low-Sr, low-Ba region reflects the classical r-enrichment by
massive SN.  However, in this Figure we can also see a sort of branch or plume
of stars having a high Sr content, and proportionally less Ba. This group of
stars seems to lie outside the general trend.  If confirmed, this could imply
that a unique site of r-process production might not be not sufficient, and we
would require an additional nucleosynthesis mechanism able to provide almost
exclusively light n-capture elements (such as Sr) with only a small amount of
heavier n-capture elements (like Ba). This idea is not new (Wasserburg \& Qian
2000), but it will require a much larger sample of stars, analyzed in a
consistent and homogeneous manner, in order to be tested.

\subsection{A ``Big" Sample \label{bigsample} }

The right panels in Figure~\ref{fig_alpha1}, Figure~\ref{fig_alpha2},
Figure~\ref{fig_ironpeak} and Figure~\ref{fig_heavy}, as well as the lower
panel in Figure~\ref{fig_al} show all the data displayed in the left panels as
small filled circles, regardless of their source.  Also, additional data were
added from other studies which were not included in the left panels to maintain
the clarity  and avoid overcrowding of the latter.  Following
NRB2001, we add into these right panels data from Gilroy \etal\ (1988), Nissen
\& Schuster (1997), Carney \etal\ (1997) and 
Stephens (1999)\footnote{NRB2001 made an effort in their paper to
bring onto a homogeneous system the previous data from the literature
that they used to build up their comparison sample. However, unless the
measured EWs are re-analyzed in a same fashion, using the same model atmospheres
and the same procedure to derive atmospheric parameters, we 
cannot exclude the possibility that
residual systematic (small) offsets are still present among 
different samples}.  Superimposed in
these panels are lines indicating abundance trends determined with robust
statistical tools.  The summary lines in these figures are described in detail
in NRB2001; we present a brief description below.

The summary lines are robust locally weighted regression lines (abbreviated as
{\it loess} lines) described by Cleveland (1979, 1994)\footnote{The source
  code for {\it loess} regression can be obtained from
  http://www.astro.psu/edu/statcodes/sc\_correlregr.html. It is also
  available as a regression option in many commercially available statistical
  packages.}  , and determined as follows.  First, average values of each
abundance ratio were obtained.  Next, we obtained three summary lines -- the
central {\it loess} line (CLL), the lower {\it loess} line (LLL), and the
upper {\it loess} line (ULL), as a function of the [Fe/H] values.  The CLL is
defined as the {\it loess} line when all the data are considered, and provides
our best estimate of the general trend of the elemental ratios at a given
[Fe/H].  Next, residuals about the CLL were obtained, and separated into those
above (positive residuals) and below (negative residuals) this line.  The LLL
is defined as the {\it loess} line for the negative residuals as a function of
[Fe/H].  The ULL is defined as the {\it loess} line for the positive residuals
as a function of [Fe/H].  If the data are scattered about the CLL according to
a normal distribution, the LLL and ULL are estimates of the true quartiles.
The {\it loess} lines remain sensitive to local variations without being
unduly influenced by outliers.  Furthermore, they are able to better handle
the endpoints of the data sets than the more commonly used median lines.  In
each subpanel the CLL is flanked by the ULL and CLL.

In order to quantify the abundance scatter in these diagrams we compute the
scale\footnote {The scale matches the dispersion for a normal distribution.} of
the data for each elemental ratio, making use of the CLL obtained above, and
consider the complete set of residuals in the ordinate of each data point about
the trend.  In Table~\ref{t:scat} we summarize robust estimates of the scale of
these residuals over several ranges in [Fe/H], using the biweight estimator of
scale, $S_{BI}$, described by Beers, Flynn, \& Gebhardt (1990).  The first
column of the table lists the abundance ranges considered.  In setting these
ranges, we sought to maintain a minimum bin population of $N = 15-20$.  The
second column lists the mean [Fe/H] of the stars in the listed abundance
interval, and the third lists the numbers of stars contained in that interval.
The fourth column lists $S_{BI}$, along with error bars obtained by analysis of
1000 bootstrap resamples of the data in the bin.  These errors are useful for
assessing the significance of the difference between the scales of the data
from bin to bin.  Note that, with the inclusion of the newly measured data from
our present paper, as well as from other recent sources, we are able to provide
scatter estimates for a somewhat finer grid, extending to lower metallicities,
than was presented by NRB2001.

\subsection{The Scatter in the [Mg/Fe] Ratio and its Interpretation}

The intrinsic scatter in the element-to-element ratios at various metallicities
contains valuable information about the typical size of the clouds undergoing
independent chemical evolution in the early epochs of halo formation, as well
as on the typical number of supernovae (SN) that polluted such clouds. For the
purpose of this discussion we assume that only core-collapse SN are important
contributors to element production in this early phase of the Galaxy. We will
return later to this point, to briefly comment on the possible impact of
nucleosynthesis from Very Massive Stars (VMS).

The most interesting elements in the present context are Fe (assumed to be
representative of the abundance of Fe-peak elements), and the
$\alpha-$elements, because the ratio in an EMP star
of the abundance of Fe to that of the
$\alpha-$elements is expected to be quite sensitive to the original mass of the
SN. Important information is also provided by other elements (such as those
produced by rapid n-capture): these have been considered by other authors (see
e.g. McWilliam 1997; Qian \& Wasserburg 2001a, 2001b, 2001c). However, it is
possible that only SN with progenitors in a restricted mass range have
contributed significantly to the production of many of these other elements,
and the exact mass ranges are not known at present.

We  concentrate here on the Mg/Fe ratio, since this
is available for a large number of stars with [Fe/H]$<-2$, is not overly
sensitive to the details of the abundance analysis 
(a major concern for O: see e.g.
Asplund \& Garcia Perez, 2001), and is less sensitive to details of
nucleosynthesis than ratios involving other elements, e.g. Ca and Ti. 
From the numbers given in the previous subsection, 
we note that the r.m.s. scatter for the ratio
[Mg/Fe] over the available sample of EMP stars is
$0.19\pm 0.03$\ near [Fe/H]=$-3$, and $0.16\pm 0.03$\ near [Fe/H]=$-2$.
Part of this scatter must arise from problems in the observations and
analysis, rather than being intrinsic. This is certainly the case for 
another element of interest, Si,
for which the analogous values are $0.37\pm 0.05$\
and $0.13\pm 0.02$. The large scatter at very low metallicities 
of the [Si/Fe] ratios can be
attributed to the fact that Si abundances in EMP stars are 
usually derived from a
single line (at 3905\,{\AA}), and are thus very uncertain. However, in the case
of Mg, the observational constraints 
are less severe as there are several clean lines of Mg I which are strong
enough to be detectable in EMP stars.  
Since we are only interested here in order
of magnitude estimates, we will ignore any non-cosmic scatter
(e.g. due to use of sub-samples from other studies)
in the Fe and Mg abundances as we anticipate that measured values for 
the Mg/Fe scatter are
already quite small with respect to expectations, so that any further
reduction would strengthen our conclusion.

Comprehensive treatments of the scatter in element-to-element abundances among
EMP stars have been recently produced by various authors (see, e.g., Argast
\etal\ 2000; Karlsson \& Gustafsson 2001, and references therein). Models
that take into account the stochastic effect of pollution from individual SN,
as well as the lifetime of their progenitors, have been developed.  Such models
allow a detailed description of the interplay between stellar evolutionary
times and the time required for a complete mixing of a suitable fragment of
the original halo. However, such modelling requires various input quantities
(e.g. yields) that are not well known at present, so that conclusions can only
be reached at order of magnitude levels. Furthermore, it is not entirely clear
that the models adequately reproduce the mechanisms of star formation and mixing
within the ISM (e.g., stars are considered to form individually, rather than in
clusters). In the following, we offer a much simpler approach, that allows
easy insight into some important issues, while still mantaining order of
magnitude accuracy. We invite
the reader to consider all the following results as very preliminary and model
dependent.  Once better understanding of
the basic physics becomes available, complex models such as those considered by
Argast \etal\ (2000) or Karlsson and Gustaffson (2001) must be considered, with
possibly an even more elaborate formulation for the star formation process and
for mixing within the ISM.

The essential ideas of our approach are to consider the early Galaxy as made
of several independent clouds, and treat each cloud as a closed box\footnote{In
chemical evolution models terminology, a "closed box model" is a model where
there is no exchange of matter with external components (that is, neither
infall or outflow of matter). Closed box models are very simple (see e.g.
Pagel 1997). An important property of the closed box models is the simple
relation existing between the concentration of metal $i$\ in the gas $z_i$\,
and the fraction of matter still in gas form $g$:
\begin{equation}
z_i = y_i\, \ln g,
\end{equation}
where $y_i$ is the yield of the metal $i$\ through production in
stellar interiors.} undergoing
its own chemical evolution. We further assume that the ISM of each
protogalactic cloud, from which the EMP stars we currently observe formed, was metal
enriched by the ejecta of SN produced by a single generation of progenitors.
Of course, this is a very schematic approach, but we think still useful; it
might correspond to a picture where a small star cluster/association begins
forming within a primordial cloud (still with zero metals). The winds and
SN-ejecta from its most massive stars pollute the remaining part of the cloud,
from which a second generation of stars (those we currently observe) form.
Effective mixing due to turbulence is assumed to maintain chemical homogeneity
of the cloud (note that relaxing this condition would increase the expected
scatter in the element-to-element ratios). In a closed box model, the metal
abundance of the ISM is set by the yields and by the fraction of gas still
remaining. Hence, once the yields are known (from SN models and from
assumption of an initial mass function, IMF), the fraction of gas remaining is
unequivocably determined. The next step is to insert discreteness, that is, a
finite number of SN. If SN yields for individual elements are not constant,
and depend, for example, on the initial stellar mass, we should expect a
scatter in element-to-element abundances obtained from different clouds,
depending on the particular set of SN that exploded in a given cloud. Due to
Poisson statistics, we expect that the scatter will be a function of the
actual number of SN contributing to typical clouds. Since the number of SN for
a given total mass is fixed by the IMF, it is possible to normalize the total
mass in stars, and from the fraction of gas (given by the overall
metallicity), to derive the total mass of the cloud.

The above model can easily be simulated by using an appropriate Monte-Carlo
code. Essentially, we need to assume an IMF (here, we used the Miller \& Scalo
1979 IMF; note that the low mass cutoff of the IMF is not critical here; it only
affects the number of low mass stars of the very first generation expected
to still exist on the lower main sequence at
present), and yields for different elements as a function of mass. These are
by far the most uncertain quantities at present; the yield of Fe is
particularly uncertain as it strongly depends on the assumed mass cut in the
SN model, a poorly known quantity. Current SN models are unable to provide
firm values, due to their failure to naturally produce SN explosions (e.g.,
Woosley \& Weaver 1995). The very sparse observational data suggest
that the Fe mass produced remains fairly constant with increasing progenitor 
mass, with
considerable scatter (e.g. Iwamoto \etal\ 1998; Turatto \etal\ 1998; see
the discussion in Nakamura \etal\ 1999, and in particular their Figure 14,
which shows the Fe mass produced in a number of SN as a function of
progenitor mass).  The Fe yields may even be a function of 
stellar properties other than the initial mass.

In our model, we use two sets of SN yields,
those adopted by Tsujimoto \etal\ (1995) and those
given by case c of Woosley \& Weaver (1995). These particular yield predictions
were selected because they provide rather small changes of the abundance ratio
Fe/Mg in the SN ejecta with stellar masses (and thus would
be expected to agree fairly well
with the observational results mentioned above); they then predict a smaller
cloud-to-cloud scatter in the expected abundances for a given number of
polluting SN, with respect to other yield predictions (like e.g. case a and b
of Woosley \& Weaver). These models thus permit a smaller
number of SN to match the observed scatter than do other
nucleosynthesis predictions: adoption of different recipes would,
in general, lead to a larger predicted
number of SN contributing, strengthening our conclusions.

According to Tsujimoto \etal\, the Fe yield decreases by only a factor of two
over a progenitor mass range from 13 to $70~M\subsun$ (we adopted an upper
mass limit of $100~M\subsun$ in this case). According to Woosley \& Weaver
case c, it rises by about a factor of five between 11 and $40~M\subsun$ (in
this case the upper mass limit was set at $50~M\subsun$). These two predictions
roughly bracket the available data (see Nakamura \etal\ 1999). Using these yield
predictions, the scatter in the yield ratios between Fe and $\alpha-$elements
is mainly due to the large increase in the production of the latter with
masses over the same range (a factor between 50 and 100 for O and Mg), a
reasonably sound prediction of the pre-supernova models. Hence, according to
these models, a large fraction of O and Mg are produced by few SN with very
massive progenitors, while Fe is mainly produced by the less massive SN
(because they are much more numerous). A random extraction over a small number
of SN may then easily produce a large scatter in the O/Fe and Mg/Fe ratios.

In our simulations we considered cases with initial masses in stars of 10$^3$\ 
and 10$^4~M\subsun$\ respectively (these may be interpreted as the masses of
the first forming cluster/association)\footnote{These values were considered
  because the resulting expected r.m.s. scatter brackets the observed values
  for Mg/Fe.}. Assuming that all stars with masses larger than 10~M\subsun\ 
explode as SN, we expect respectively 3.4 and 34 SN in the two cases, with our
choice of IMF. We then carried out 10$^5$\ and 10$^4$\ trials respectively for
the two cases of SN distributed according to the Scalo IMF, and computed the
total masses of Fe and Mg produced by these SN for each trial.  The rms
scatter of the [Mg/Fe] ratios from these sets are 0.43 and 0.14 dex
respectively in the two cases when the Tsujimoto \etal\ yields were used (the
scaling between these two values agrees well with the larger number of SN of
the second case). As expected, the scatter is smaller when the Woosley \&
Weaver case c yields are used: in these cases, we obtain r.m.s. values of 0.29
and 0.08 dex, respectively.

When we compare the predictions of this simple model with observations, we
derive the typical number of SN contributing to the ISM from which 
the observed stars
formed as $\sim 18$\ at [Fe/H]=$-3$, and $\sim 26$\ at [Fe/H]=$-2$, when the
Tsujimoto \etal\ yields are used. The corresponding values when Woosley \&
Weaver case c yields are used are 7 and 10. 
The mass in Fe produced by $\sim 20$~SN randomly extracted using the Scalo IMF
and the Tsujimoto \etal\ yields is $\sim 2~M\subsun$; in the case of 7 SN with
Woosley \& Weaver case c models, it is $\sim 0.2~M\subsun$. If the
SN ejecta are used to
raise the metallicity of a cloud up to [Fe/H]=$-3$, the total original
(baryonic) mass of each of the clouds is $\sim 10^6~M\subsun$\ in the first
case, and ten times less in the second one. These values agree fairly well with
the expected Jeans mass at this epoch, and with typical values for (present
day) dwarf spheroidals and globular clusters. 
We intend to explore the connection between globular clusters and field
stars in a future paper.
With the caveats discussed above
kept in mind, the value we have derived might be considered to be the
characteristic mass of protogalactic fragments. Note that the typical mass of
the hypothetical primordial clusters/associations are in all cases of the order
of a few thousand $M\subsun$; it is difficult for such small objects (if
they really existed) to have remained bound after the violent mass loss that
probably occurred during their early phases. 

The typical number of SN contributing to metal enrichment of EMP stars we
obtain seems quite large, with respect to the usual assumption that metals in
these stars were produced out of material polluted by ejecta from very few,
possibly only one SN. Note that the requirement of a very small number of SN
(the basic ingredient of stochastic models of metal enrichment) mainly comes
from the large observed scatter in elements produced by neutron-capture
processes. Our result is a consequence of the relatively small scatter
observed for the [Mg/Fe] ratio, and of the assumptions about the SN yields and
the IMF. A considerable reduction in the scatter predicted by these models
could probably be obtained by limiting the mass range of the IMF, because SN
models predict a dependence of the Mg/Fe ratio on progenitor mass. A smaller
mass range could in principle be understood if e.g. star formation in the
remaining part of the cloud was very fast, and only the most massive stars of
the first generation could evolve rapidly enough to contribute to the
pollution of the ISM. However, this does not seem a palatable explanation for
various reasons:
\begin{itemize}
\item The naive expectation is that the mass range should be limited to the
  most massive SN. On the basis of nucleosyntheis predictions, we expect that
  these SN would produce a [Mg/Fe] much larger than the average over the
  entire relevant mass range. This average value should be more appropriate 
  for the
  most metal-rich halo and thick disk stars. We would then expect to observe
  in EMP stars a Mg/Fe ratio much larger than in most metal-rich halo stars:
  however, Figure~\ref{fig_alpha1} shows that the Mg/Fe ratio in EMP stars is
  similar to that observed in more metal-rich objects.
\item The previous consideration forces us to limit the mass range to those SN
  that produce [Mg/Fe] values close to the average, that is SN of about 20
  M\subsun. We are not aware of any simple physical explanation favoring this
  mass range.
\item In order not to further increase the predicted scatter, masses for
  different clouds should be assumed to be similar. This might possibly be
  justified because the mass of these clouds is indeed of the order of
  magnitude of the Jeans mass; however it should be recalled that for the
  adopted IMF the number of SN predicted for a typical Jeans mass is $>10$.
\item Finally, even more detailed models such as those of Argast \etal\ (2000)
  show a scatter in the [Mg/Fe] ratios much larger than given by observations.
  This demonstrates that our result is robust, and is not due to the way we
  modeled the early process of chemical enrichment, but rather to the
  assumptions made about the details of the closed boxes, yields, and the IMF.
\end{itemize}

There are at least two alternative scenarios to explain the surprisingly small
observed scatter in the [Mg/Fe] ratio among EMP stars, without invoking
changes in the yields, and still saving the basic concept of the stochastic
model.  First, it might be assumed that a generation of very massive ($> 100
M\subsun$) stars (VMS) polluted the medium before the formation of the EMP
stars (Qian \& Wasserburg 2001a,b).  Abel, Bryan \& Norman (2002) present a
fully self-consistent 3-d hydrodynamical simulation of the formation of one of
the first stars in the Universe.  This is an interesting hypothesis, because
the introduction of a new actor only playing at very low metallicities (and
without a corresponding low mass population, given the peculiar top-heavy IMF
that should be appropriate at zero metals; Oh \etal\ 2001) might help to
explain the changes of trend$/$scatter in several abundance ratios observed in
EMP stars, without perhaps violating other observational contraints. A full
discussion is beyond the present paper; however, we wish to note here a few
difficulties within this scenario that should be resolved before this
hypothesis can be definitively adopted:
\begin{itemize}
\item Current nucleosynthesis predictions for pair instability SN resulting
  from the evolution of VMS (Heger \& Woosley 2001) do not match well the
  observed abundance patterns in EMP stars (see also Umeda \& Nomoto 2002).
  Leaving aside the more uncertain aspects, such as the normalization of the
  production of Fe-peak to $\alpha-$elements, the main difficulties concern
  the large predicted overproduction of Mg and Si with respect to O, the
  odd-even pattern for Fe-peak elements (e.g., we expect a solar Cr/Fe ratio,
  at variance with observations), and the absence of production of elements
  heavier than Ni (such as Zn, which is clearly overabundant in most
  metal-poor stars).  As discussed by Umeda \& Nomoto, observations concerning
  all these features are better explained by nucleosynthesis by the most
  massive core collapse SN. These difficulties lead Qian \& Wasserburg to
  abandon theoretical predictions for nucleosynthesis yields for VMS, and to
  rely instead on empirical estimates, based on the observed abundance pattern
  in EMP stars themselves (Qian \& Wasserburg, 2001c).
\item As noticed by Qian \& Wasserburg (2001c), the concern about the small
  observed scatter in some abundance ratios is even larger when this scenario
  is adopted. In order to overcome it, either a very narrow mass range must be
  adopted for VMS, or the number of VMS required should be large (the same
  conclusion we were forced to when considering type II SNe). Given the much
  larger yields per explosion in the case of VMS, this last solution would
  create difficulties with the total observed amount of Mg and Si observed in
  EMP stars: in fact a single VMS should produce enough Mg and Si to justify
  the total mass of Mg and Si present in EMP stars over the whole Galaxy (if
  the yields by Heger and Woosley 2001 are used), and it seems difficult to
  conceive that this material could be distributed over such a large volume 
  (see however Abel, Bryan \& Norman 2002).
\end{itemize}

Alternatively, we may relax the closed box approximation. In fact, there
probably was some mass exchange between different fragments; in particular, we
should expect selective mass-loss through metal-rich winds from each fragment
(probably the fragments were not able to retain within themselves all the
ejecta of the SN).  Furthermore, different fragments might have formed stars
at different epochs. It is then easy to imagine that a 
fraction of
the very metal-poor fragments was probably polluted not by the ejecta of SN
explosions of massive stars (or even VMS) that formed within the fragments, but
rather from material coming from other (probably much more massive) fragments
that had a faster evolution. It is also possible to speculate that these larger
fragments having a faster evolution do not properly belong to the halo, but
rather to the material that subsequently 
formed the bulge or even part of the thick
disk. The low metallicity of some of the EMP stars would in this case 
be simply the result of dilution.

Another way of stating the above is that the number of SN we deduce represents
some average over a wide distribution of values: in some cases only one SN
contributed (this is required to produce e.g. the stars with very large
abundances of r-process elements), while in other cases there were many of
them ($>10$, and possibly even $>100$). This substantially  modifies  the
distribution of [Mg/Fe] values from that given by a population of clouds all
of which have same size.

If a large enough number of stars are observed with sufficiently high-quality
data (and treated with a uniform abundance analysis), we might even try to
verify this scenario. The distribution of [Mg/Fe] predicted from this
scenario should be very different from that given by one composed of closed
boxes: there should be a compact core, due to stars formed either in very
massive clouds, or in clouds that received their metals from outside, and a
much broader ``haze'' due to those stars formed in smaller self-polluted
clouds.  Clearly, much larger samples of EMP stars and better calculations of
SN yields are needed for significant progress in constraining these ideas.

\subsection{The CH-star HE~0024$-$2523: A Very Peculiar Object}

One of the stars in our sample falls in the region of the [Ba/Fe] vs. [Fe/H]
plane populated by objects with very high Ba abundances.  We note that this
star (HE~0024$-$2523) clearly shows a G-band of CH between 4300 and 4325\,{\AA}\
(Q-branch).  The G-band is much more intense than expected, given the rather
low metallicity of this star ([Fe/H]$=-2.63$) and its relatively
high \teff. A preliminary comparison with
synthetic spectra for the region around the band-head supports a value of the
[C/Fe] ratio of about +2.2.  This may be another case of the so-called CH-stars (or
Ba-stars), often found to be in binary systems (see McClure 1997 and McClure 
\& Woodsworth 1990). In these systems, the companion
evolves faster and passes through the thermal pulsing AGB phase, producing a
large amount of C and s-elements. In the subsequent evolution of the system,
this enriched material is ejected onto the star we presently observe.  CH stars
are expected to occur frequently among metal-poor dwarfs (and subgiants), given
the small amount of material, processed within the thermally pulsing star,
required to significantly pollute the outer convective envelope of metal-poor
turn-off stars.

In HE~0024$-$2523, europium (an almost totally r-process element) is not
observed (we checked the 4129.7\,{\AA} line), Sr is enhanced ([Sr/Fe]$\sim 0.5$),
while Ba is very strong ([Ba/Fe]$\sim 1.7$): this supports the identification
of this star as a CH star.

Moreover, in very metal-poor stars (such as the one under scrutiny) there is a
paucity of seeds able to capture neutrons along the s-chain. As a consequence,
the neutrons accumulate, and heavier and heavier elements are built, so that
the whole synthesis is shifted toward the heaviest nuclei compared to solar
abundances.  We can then expect a large abundance of lead in this star
(Travaglio \etal\ 2001).

The only Pb line available in our spectrum is the line at 4057.815\,\AA. This
line is heavily contaminated by a nearby CH-line (see also Aoki \etal\ 2000). 
We thus compared the observed
profile with the result of a full spectral synethesis (see
Figure~\ref{fig_pb}). The line list used in these computations was originally
taken from Kurucz (1995) and adjusted in order to get a good fit to the solar
flux spectrum of Kurucz \etal\ (1984). To improve the S/N, the observed spectrum
was smoothed with a Gaussian having FWHM=0.06\,{\AA} ; this is much less than the
intrinsic line width in the spectrum of this star, since the lines are
significantly broadened by rotation (see Paper I). The lead line appears to be
detected, supporting an enormous Pb abundance ([Pb/Fe]$\sim 3.2$).

There is no doubt that this is a very interesting object, and we have already
re-observed this star to confirm the peculiar abundance pattern shown by this
class of stars (and in particular the enormous Pb abundance) and to search for
the expected radial velocity variations associated with binarity.  A detailed
analysis of this star will be discussed in a separate paper (Gratton {\etal},
in preparation).

\section{Summary}

We have presented a detailed abundance analysis for eight stars
(seven expected to be near the main sequence turnoff, and one probable giant)
selected as extremely metal-poor candidates from the Hamburg/ESO Survey. For
comparison, we analyzed also three stars (two giants and one dwarf) from the HK
survey, and three additional very bright metal-poor stars.  With this work, we
have doubled the number of extremely metal poor stars ([Fe/H] $\le -3.0$ dex)
with abundance analyses based on high precision, high-spectral-resolution data.

Since we have utilized stellar parameters determined independently of the
spectra, the analyses of the spectra themselves yield parameters that can be
used as diagnostics to test the validity of the atmospheric parameters assigned
in Paper I; these appear to be valid to within the uncertainties given in Paper
I.  We looked for evidence of departures from LTE for Fe, and did not find any,
with upper limits at a level of 0.1 to 0.2 dex.

We studied the $\alpha-$elements Mg, Si, Ca, Ti; the light element Al; the
iron-peak elements Sc, Cr, Mn; and the neutron-capture elements Sr and Ba.  The
first key result is that our sample of HES EMP candidates contains three stars
with precision Fe abundances from the present high-dispersion analysis with
[Fe/H] $\le -3.0$ dex, three more with [Fe/H] $\le -2.8$ dex, and the remaining
two stars are only slightly more metal rich.  Thus the chain of procedures that
led to the selection of these stars successfully produces a high fraction of
extremely metal poor stars.

In general, when we combine our sample with data from the literature, our
results support the trends of element variation [X/Fe] with decreasing [Fe/H]
found by previous investigators.  These trends appear to be the same for
dwarfs and for giants, extending even to the low metallicities studied here.

However, we are struck by the {\it lack of scatter} in abundance ratios among
most elements in these EMP stars.  While it is well known that among stars with
solar metallicity, abundance ratios are essentially constant, showing only
small trends with time, we naively expected stochastic effects of small numbers
of SN to have become imporant at the extremely low abundances we are exploring.
Much to our surpise, we find that most elements have quite a small scatter
(${\lesssim}0.1$ dex, some of which undoubtly arises from experimental
errors) in abundance [X/Fe] even at [Fe/H] $\le -3.0$ dex.  Among the elements
studied, only Sr and Ba show large scatter at a fixed (very low) Fe/H, and
perhaps we have begun to discern genuine scatter for Mn and Cr.

We discuss the implications of these results, and suggest that we are almost at
the point of having samples large enough to be able to constrain
nucleosynthesis in the early evolution of the Galaxy, and in particular, the
size of independent clouds at the time these EMP stars were formed in the
Galactic halo.  The preliminary value for the characteristic mass of
protogalactic fragments that we deduce is tantalizingly close to that of the
expected Jeans mass at this epoch, and to typical values for (present day)
dwarf spheroidals and globular clusters.

As the work of the 0Z project advances over the next few years, we may look
forward to much larger samples of EMP stars with accurate abundance
determinations becoming available.  The confrontation with the theory raised by
the small scatter in [X/Fe] seen thus far at extremely low metallicity will
become much sharper.  If these small scatters persist, the assumptions normally
adopted for the early chemical evolution of the Galaxy will have to be
re-examined, requiring improved calculations of nuclear yields, the possibility
of a previous generation of very massive stars, or mixing between
``independent'' primordial clouds.

Among such extremely metal poor stars, the pollution of a stellar atmosphere
with a relatively small amount of some heavy element (perhaps from a binary
companion) will tend to produce a detectable abundance enhancement.  As a
preview of the ``zoo'' of peculiar stars we may thus expect to find in our
future work, we note that one of the stars in our small initial sample, 
HE~0024$-$2523, was found to be a CH star, with extremely enhanced Ba and
somewhat enhanced Sr.  This  main-sequence star also shows a very large
overabundance of lead, and appears to represent the result of the s-process
chain operating in a very metal poor environment.

\appendix

\section{Comparison of Our Adopted $gf$ Values with Those of NIST}

We compare the $gf$ values we have adopted in the present work, which have
been assembled from the sources listed at the end of Table~\ref{t:ewbright},
with those of NIST.  The NIST Atomic Spectra Database Version 2.0 (release
date March 1999) (NIST Standard Reference Database No. 78, URL
http://physics.nist.gov/cgi-bin/AtData/main\_asd) provides access to
critically evaluated data on atomic energy levels, wavelengths, and transition
probabilities that are reasonably up to date.

For each ion we consider, Table~\ref{t:nist} gives the number of lines in
common with the NIST database from the set given in Table~\ref{t:ewbright} and
Table~\ref{t:ewhes}, then the mean of the values of $\Delta[gf$(NIST)
$-gf$(us)], as well as the dispersion of the differences about the mean.

In general, the results are quite satisfactory.  The mean difference ranges
from $-0.02$ dex to 0.00 dex.  For most of the ions considered here, the
dispersion of the differences is small.  However, in a few cases (Mg I, Ti II,
and Fe II) the dispersions are larger and are discussed below.

Table~\ref{t:nist} serves to remind us that $gf$ values are still uncertain and
that systematic errors of normalization exist between the results of different
teams at the level of $\sim$0.1 dex, with smaller internal uncertainties within
each dataset.

\subsection{Mg I}

The present status of the oscillator strengths for Mg I is not very good, and
oscillator strengths for Mg I lines are generally derived from theoretical
calculations.  Those used by most authors are from Froese-Fischer (1975), and
are used in the solar analysis of Lambert \& Luck (1978) as well.  However,
these are not the most recent ones.  The latest updated computations are given
by Mendoza \& Zeippen (1987), and by the Opacity Project group (=TOP, results
available through CDS).  Since the opacity project group includes Mendoza, we
view the  TOP $gf$ values for Mg I as updates of the Mendoza \& Zeippen
calculations. Values from the TOP group agree fairly well with those from
Froese-Fischer. The VALD database uses data from the Kurucz CD-ROM18.

A comparison between different sources for the relevant transitions of Mg I is
given below in Table~\ref{t:gfmg}. Kurucz $gf$'s agree well with the TOP
calculations for the triplet lines, while they are lower by 0.2 to 0.3 dex for
the singlet lines.

The reason we prefer to use Kurucz $gf$'s (those in the VALD database) is that
(in other parallel studies currently in progress) with these $gf$'s we get
better agreement among the Mg I lines in well studied moderately metal-poor
stars ($-2<$[Fe/H]$<-0.5$) with weaker Mg I lines (at 6318\,{\AA}, etc.), those
generaly used in solar abundance analyses.

\subsection{Ti II}

For Ti II, whenever possible we adopted experimental $gf$'s from Bizzarri
\etal\ (1993). These were obtained by combining branching ratios from hollow
cathode measurements with lifetimes from selective induced laser excitation.
For most lines they are accurate within about 10\%. For the remaining lines,
they were taken from Magain (1985), who discussed literature values available
at that epoch. When neither of these were available, we adopted those from
Kurucz CD-ROM18 (1995).

When compared with values from the NIST database, these $gf$'s show significant
scatter (although the average values agree). The NIST $gf$'s for Ti II are
mostly from the Martin, Wiese \& Fuhr (1983) compilation, with a few additions
from Kurucz's CD-ROM18. They do not include the most recent experimental
values. When used in our abundance analysis for the three bright comparison
stars, they produced a somewhat larger line-to-line scatter in the abundances.

Very recently (after our analysis was completed), Pickering, Thorne  \& Perez
(2001) presented a new set of experimental $gf$'s, based on new branching
ratios from hollow cathode measurements coupled with experimental lifetimes.
Their line list is more extensive, but the individual values are less accurate than
those of Bizzarri \etal, with typical accuracies of 10 to 20\%. When compared
to the set of $gf$'s adopted in our paper, the mean difference is $-0.02\pm
0.03$, with a r.m.s. value of 0.12~dex from 18 lines.  The scatter is
definitely larger when comparison is made with $gf$'s from those lines in the
NIST database included in our list ($-0.05\pm 0.04$, with an r.m.s. of
0.16~dex). We conclude that the adopted $gf$'s are to be preferred to those of
the NIST database.

\subsection{Fe II}

The line-to-line comparison shows quite a large scatter (but no zero-point
offset) between our adopted $gf$'s for Fe II and those from NIST. These
last are those from Kurucz CD-ROM18, which are from semi-empirical
calculations.  Our Fe~II $gf$'s are the average of the experimental values by
Heise \& Kock (1990) and Hannaford \etal\ (1992), of the theoretical ones by
Bi\'emont \etal\ (1991), and of solar $gf$'s from Blackwell \etal\ (1980)
(these last were increased by 0.19 dex to put them on the same scale as given by
the three other sources), save for four lines missing these data for which the
Kurucz $gf$'s were adopted. For those lines where the two values disagree, our
adopted $gf$'s are most likely more accurate that those listed in NIST.

After completing our analysis, we become aware that two more recent papers have
been published with Fe II $gf$'s: the theoretical calculations by Raassen \&
Uylings (1998), and the experimental ones by Schnabel \etal\ (1999).  These two
sources agree very well with previous determinations, but with a systematic
offset of 0.10 dex in the $gf$'s by Raassen \& Uylings. Since differences
between our adopted $gf$'s and those from these two last sources are very small
($<0.05$~dex, once Raassen \& Uylings $gf$'s are put on the same scale as those
from the other authors) save for one line\footnote{For the
Fe~I line at 4233.17\,{\AA}, Kurucz 
gives $gf = -2.00$, which we adopt in our analysis, 
while the value given by
Schnabel \etal\ is $-1.81$.}, it was not deemed necessary to repeat our
analysis.

\acknowledgements

The entire Keck/HIRES user community owes a huge debt to Jerry Nelson, Gerry
Smith, Steve Vogt, and many other people who have worked to make the Keck
Telescope and HIRES a reality, and who continue to operate and maintain the
Keck Observatory.  We are grateful to the W. M. Keck Foundation for the vision
to fund the construction of the W. M. Keck Observatory.
The authors wish to extend
special thanks to those of Hawaiian ancestry on whose sacred mountain
we are privileged to be guests.  Without their generous hospitality,
none of the observations presented herein would
have been possible.  
 
We thank the referee, Bruce Carney, for a very careful reading of the
paper, and for his constructive comments.

T.C.B acknowledges partial support for this work from grants AST 00-98508 and
AST 00-98549 from the National Science Foundation.

This research has made use of the SIMBAD database, operated at CDS, Strasbourg,
France.

\clearpage

%
%
\begin{deluxetable}{lrrrrrr}
\tablenum{1}
\tablewidth{0pt}
\scriptsize
\tablecaption{Adopted Model Atmosphere Parameters}
\tablehead{\colhead{Star ID} & \colhead{\teff} & 
\colhead{log(g)} & \colhead{[A/H]} & \colhead{$v_t$} 
& $\Delta$[Fe/H]$_{\rm II-I}$ & $\Delta(Fe~I/\chi)$ \nl
\colhead{} & \colhead{(K)} & \colhead{(dex)} & \colhead{(dex)} &
\colhead{(\kms)} & \colhead{(dex)} & \colhead{(dex/eV)}
}
\startdata
HD 140283      & 5750 &3.67 & $-$2.44 &1.10 & $-$0.025 & $-$0.064 \nl
BD 3--740      & 6355 &4.00 & $-$2.70 &1.42 & $-$0.005 & $-$0.063 \nl
G139--8        & 6200 &4.50 & $-$2.06 &1.02 & $-$0.116 & $-$0.074 \nl
BS 17447--029  & 6530 &4.40 & $-$2.93 &1.94 & $-$0.268 & $-$0.041 \nl
CS 22950--046  & 4730 &1.30 & $-$3.30 &2.02 & $-$0.170 & $-$0.072 \nl
CS 22878--101  & 4775 &1.30 & $-$3.09 &2.01 & $-$0.053 & $-$0.058 \nl
HE 2133--1426  & 6300 &4.10 & $-$2.83 &1.31 & $-$0.138 & $-$0.045 \nl
HE 2344--2800  & 6625 &4.30 & $-$2.56 &1.42 &    0.036 & $-$0.043 \nl
HE 0024--2523  & 6625 &4.30 & $-$2.65 &0.58 & $-$0.005 & $-$0.013 \nl
HE 0130--2303  & 6560 &4.30 & $-$2.96 &1.32 & $-$0.087 & $-$0.017 \nl
HE 0132--2429  & 5310 &3.40 & $-$3.59 &1.28 & $-$0.217 & $-$0.146 \nl
HE 0148--2611  & 6550 &4.30 & $-$2.98 &0.78 & $-$0.309 &    0.002 \nl
HE 0218--2738  & 6550 &4.30 & $-$3.54 &0.84 &   ...    & $-$0.077 \nl
HE 0242--0732  & 6360\tablenotemark{a}  & 4.40\tablenotemark{a}
& $-$3.21 &0.40 & $-$0.212 & $-$0.150 \nl
\enddata
\tablenotetext{a}{Values slightly modified from those of Paper~I, see
text for details.}
\label{t:atm}
\end{deluxetable}

\clearpage

\begin{deluxetable}{rrrr} 
\tablenum{8} 
\tablewidth{0pt} 
\tabcolsep 2pt 
\tablecaption{A Comparison of our Adopted $gf$ Values with Those of NIST}
\tablehead{\colhead{Ion} &   \colhead{No. Lines} &
 \colhead{$<gf$(NIST)$-gf($us)$>$} &
\colhead{~~~$\sigma$} \nl
  \colhead{} &  \colhead{} &  \colhead{(dex)} &  \colhead{(dex)} \nl
}
\startdata
Al I   &    2  &  0.00   & 0.00 \nl
Mg I   &   3\tablenotemark{a} & +0.10 & 0.17 \nl
       &  2\tablenotemark{b} & 0.00 & 0.00 \nl
Si I   &    1 &   0.00    &  ... \nl
Ca I   &    6 &   0.00    &  0.00  \nl
Sc II  &    4 &   0.01    &  0.01 \nl
Ti I &  16 & $-$0.01 & 0.02 \nl
Ti II   & 26 &  $-$0.01 & 0.14 \nl
Cr I   &  14 &  0.00 & 0.01 \nl
Mn I   &  5 &   0.00  & 0.00 \nl
Fe I & 59 & $-$0.01 & 0.07 \nl
Fe II  & 12 & $-$0.01 & 0.11 \nl
Sr II  &   2   &  $-$0.02  &  0.00 \nl
Ba II  &   1   &  0.00  &  ... \nl
\enddata
\tablenotetext{a}{There are two additional observed lines of Mg I
for which there is no lifetime in either the current NIST database or
in Weise, Smith \& Miles (1966).  The NIST team is currently
updating the entries for this ion.}
\tablenotetext{b}{Here we omit the 4703 \AA\ line of Mg I.}
\label{t:nist}
\end{deluxetable}

%
%
\begin{deluxetable}{rrrrrrr} 
\tablenum{9} 
\tablewidth{0pt} 
\tabcolsep 2pt 
\scriptsize 
\tablecaption{Comparison of Mg I Transition Probabilities}
\tablehead{
Source &\colhead{FF75} 
& \colhead{TOP}
& \colhead{MZ}
&\colhead{VALD=Kurucz} 
& \colhead {CD$-$ROM18}
& \colhead {Adopted}
} 
\startdata 
Singlet transitions:
\nl
3p1P$-$4d1D 
& 5528
& $-$0.35
& $-$0.40
& $-$0.489
& $-$0.620
\nl
$-$5d1D 
&4703
 & $-$0.38
& $-$0.42
& $-$0.462
& $-$0.666
& $-$0.67
\nl
 $-$7d1D
& 4167
& $-$0.71
& $-$0.75
& $-$0.757
& $-$1.004
& $-$1.00
\nl 
$-$8d1D
& 4057
& $-$0.89
& $-0.91$ 
& ... 
& $-$1.201
& $-$1.20
\nl
3p1P$-$5s1S
& 5711
&&$-$1.75
& $-$1.724
& $-$1.833
\nl
 $-$6s1S
& 4730
&& $-$2.34
& $-$2.268
& $-$2.523  \nl
Triplet transitions:
\nl
3p3P$-$4s3S
& 5178
&& 0.09
& $-$0.018
& 0.061
&  \nl
 & ~~~(total) \nl
&5167
&& $-$1.00
& $-$1.030 
\nl
&5172
&& $-$0.37
& $-$0.402
& & $-$0.38
\nl
 &5183
&& $-$0.15
& $-$0.180
&& $-$0.16
\nl
\enddata 
\label{t:gfmg}
\end{deluxetable}

\clearpage

%
%

%


%


%

%

\begin{figure}
\epsscale{1.0}
\plotone{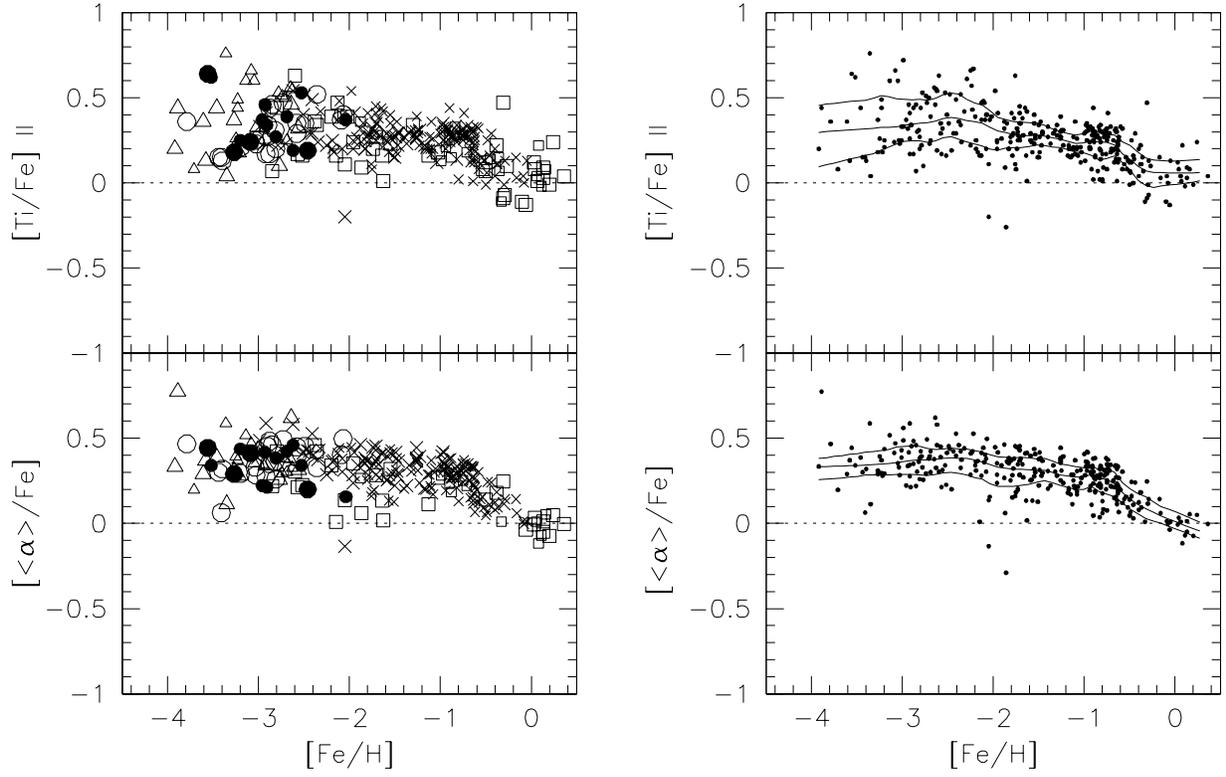}
\caption[ecarretta_fig07.ps]{The same as in Figure~\ref{fig_alpha1}, but for Ti II abundances 
and the average of the $\alpha-$elements, 
constructed from 
the abundance ratios of
the best observed ions: Mg I, Ca I and Ti II.
\label{fig_alpha2}}
\end{figure}
%


\begin{figure}
\epsscale{1.0}
\plotone{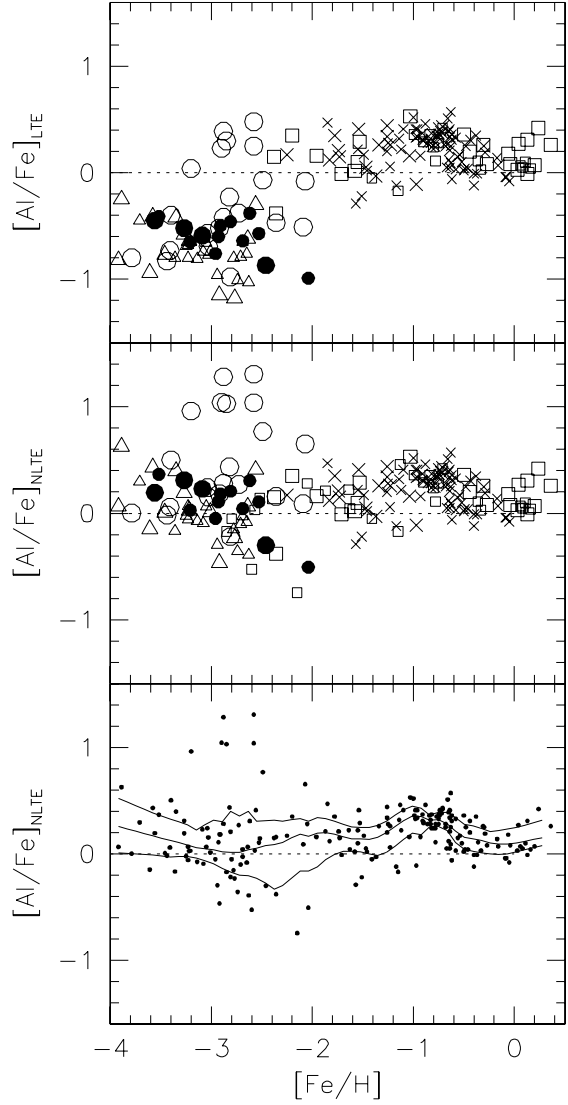}
\caption[ecarretta_fig08.ps]{Run of [Al/Fe] ratios as a function of 
metallicity. Symbols are as in Figure~\ref{fig_alpha1}. Upper panel: [Al/Fe] ratios derived in
the LTE assumption; central panel: [Al/Fe] ratios with corrections for
departures from LTE, following prescriptions by Baum\"uller \& Gehren (1997) 
(see text,
Section 5.2). In the lower panel are shown the abundance ratios [Al/Fe]
corrected for NLTE, with  the {\it loess} summary lines superimposed.
\label{fig_al}}
\end{figure}
%


\begin{figure}
\epsscale{1.0}
\plotone{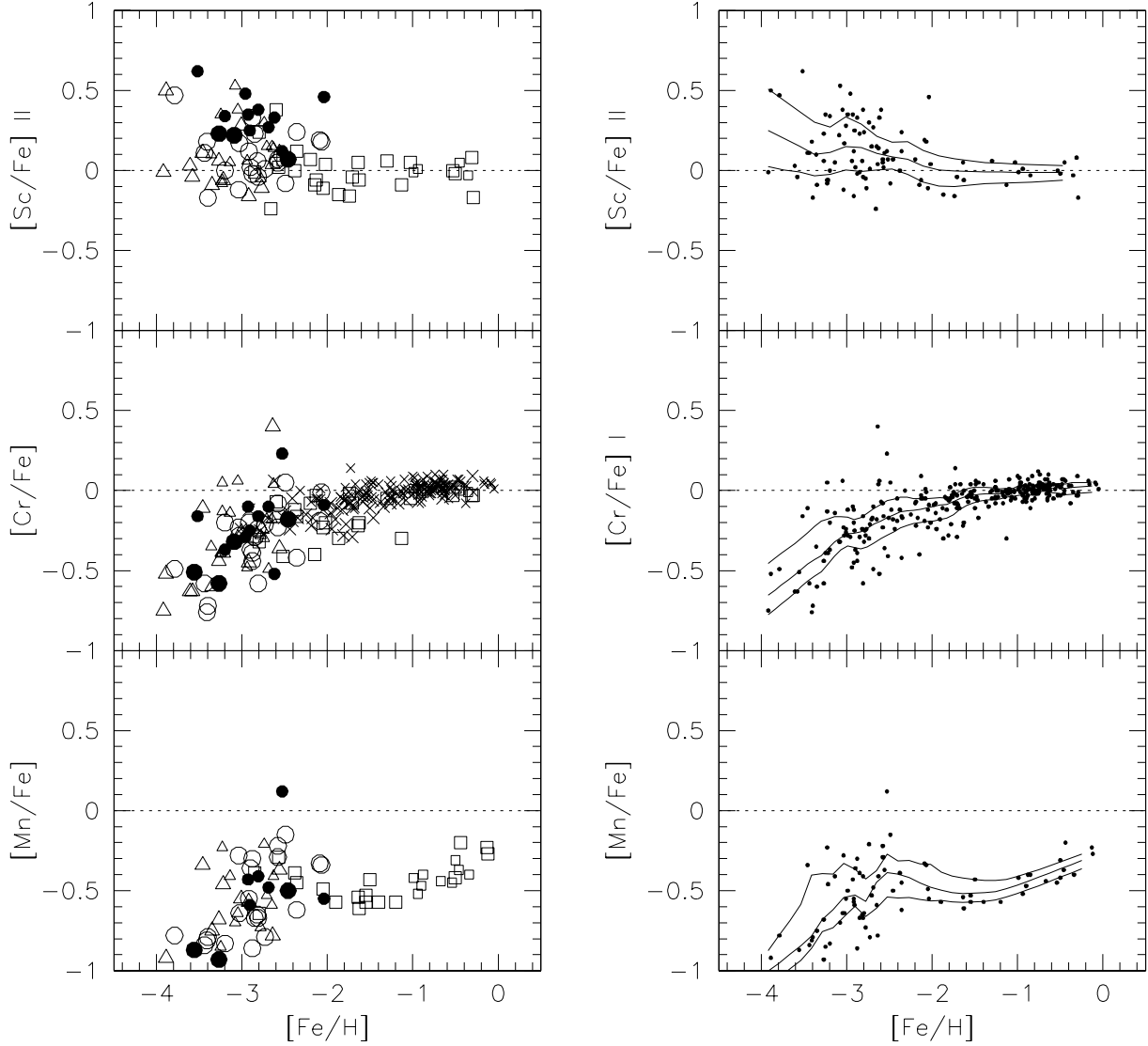}
\caption[ecarretta_fig09.ps]{Abundances of iron-peak elements in program stars as a
function of [Fe/H], as compared with data from previous studies, as in Figure~\ref{fig_alpha1}.
The left panels 
allow one to differentiate among datasets, while the right panels, with
summary lines superimposed, plot all data with a single symbol
irrespective of their source.
\label{fig_ironpeak}}
\end{figure}
%

%
\begin{figure}
\epsscale{1.0}
\plotone{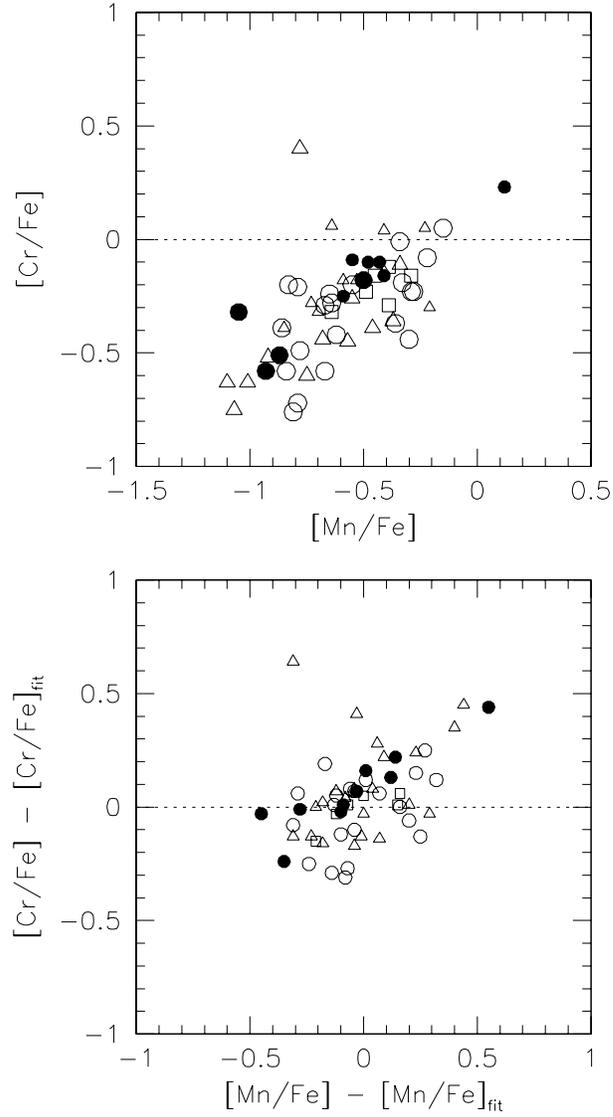}
\caption[ecarretta_fig10.ps]{
Run of the [Cr/Fe] abundance ratio as a function of [Mn/Fe] for program stars
and for  data from the literature (upper panel). Symbols are as in
Figure~\ref{fig_alpha1}. Only stars with [Fe/H$<-2$ are plotted. Lower panel: 
residual of [Cr/Fe] and [Mn/Fe] ratios with respect to linear regressions
obtained from [Cr/Fe] vs. [Fe/H] and [Mn/Fe] vs. [Fe/H] diagrams for stars with
[Fe/H]$<-2$.
\label{fig_crmn}}
\end{figure}
%

%
\begin{figure}
\epsscale{1.0}
\plotone{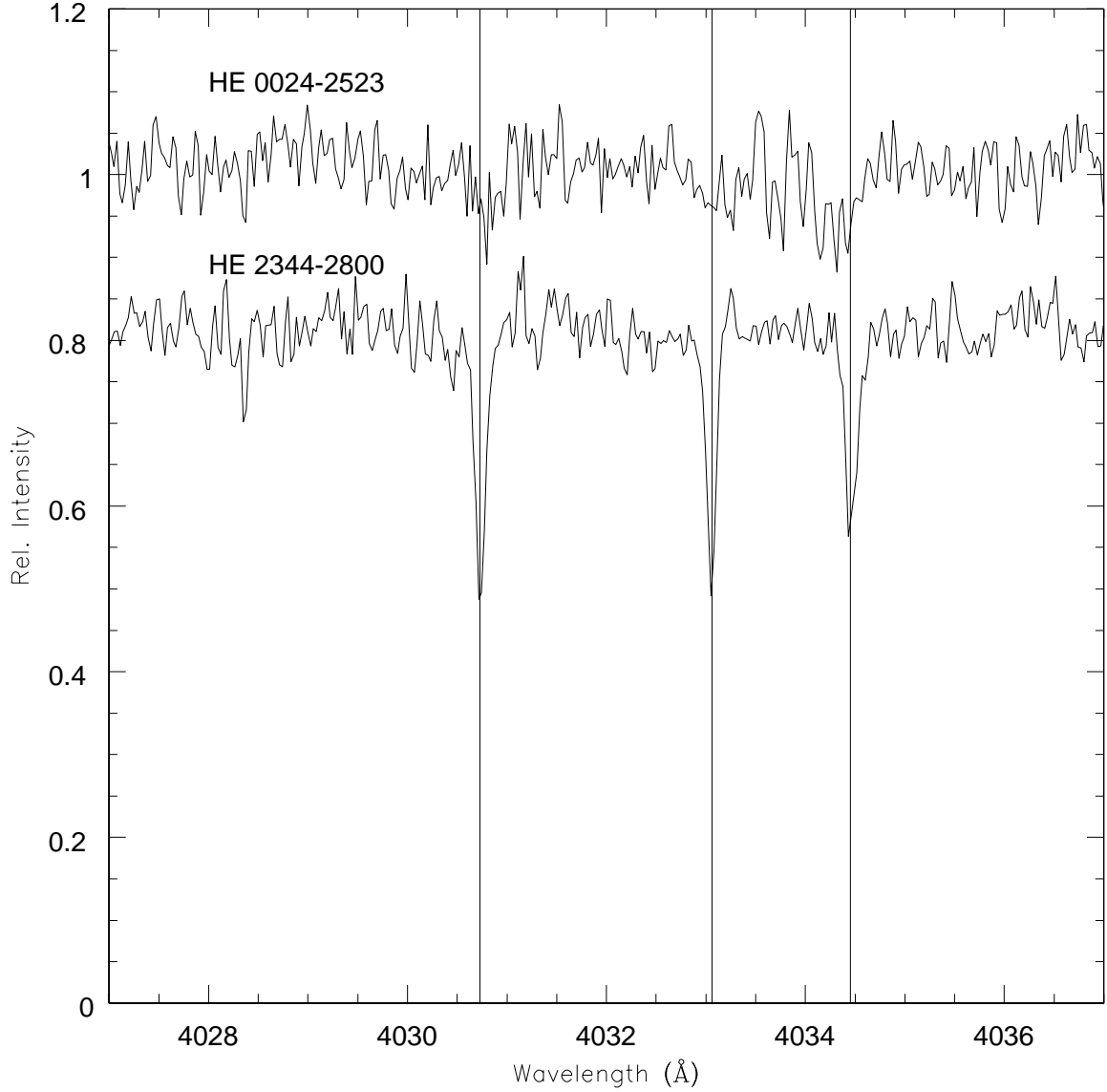}
\caption[ecarretta_fig11.ps]{
Observed spectra of the program stars HE~0024$-$2523 and HE~2344$-$2800 in the
region around the Mn I lines at 4030.75, 4033.06 and 4034.48\,{\AA} (indicated by
vertical solid lines). The spectrum of HE~2344$-$2800 has been arbitrarily 
shifted in intensity for purposes of clarity.
Note that HE~2344$-$2800 is the object with the anomalously large Mn
abundance in Figure~\ref{fig_crmn}.
\label{fig_permn}}
\end{figure}
%


\begin{figure}
\epsscale{1.0}
\plotone{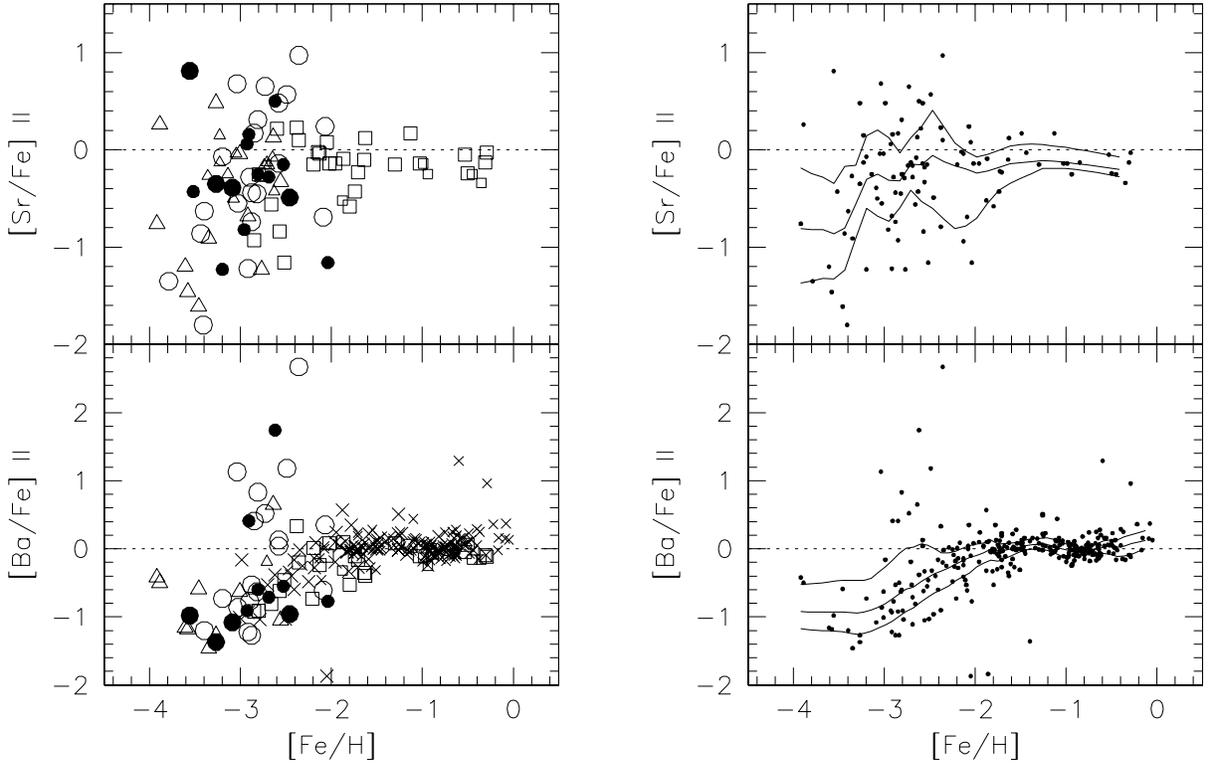}
\caption[ecarretta_fig12.ps]{Comparison of the run of element ratios of 
the n-capture elements
Sr (upper panels) and Ba (lower panels) as a function of 
[Fe/H] in the present work and other previous studies; the
symbols have the same meaning as in
previous figures. 
\label{fig_heavy}}
\end{figure}


\begin{figure}
\epsscale{1.0}
\plotone{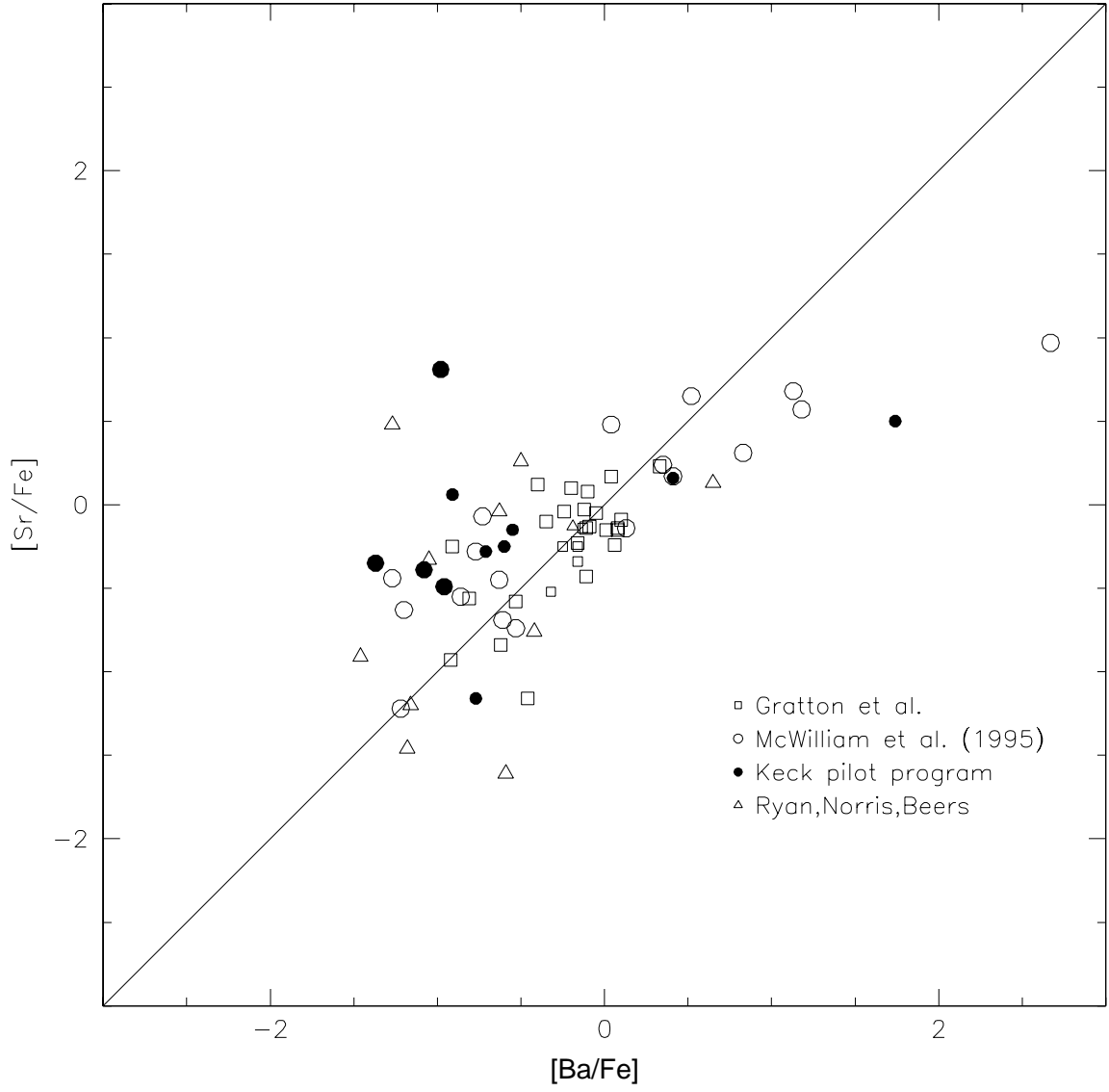}
\caption[ecarretta_fig13.ps]{The light n-capture element [Sr/Fe] as a function of the
heavier n-capture element [Ba/Fe]. 
\label{fig_srba}}
\end{figure}

%
\begin{figure}
\epsscale{1.0}
\plotone{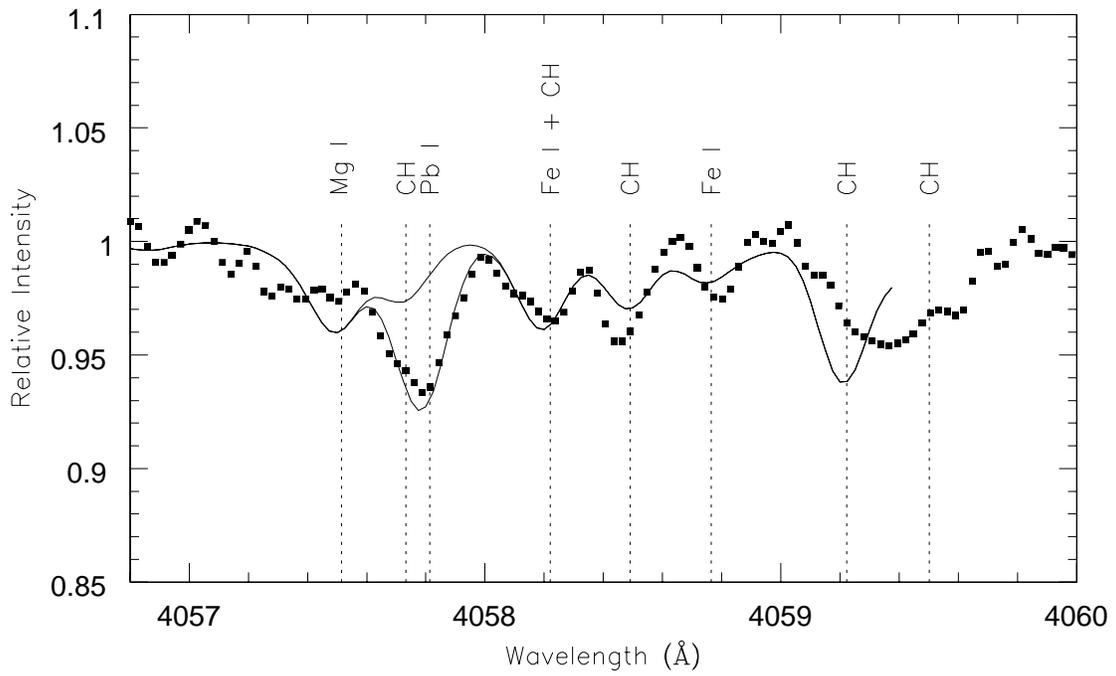}
\caption[ecarretta_fig14.ps]{A comparison of
the observed 
spectrum of HE~0024$-$2523 
in the region of the Pb line at 4057\,{\AA} and 
synthetic spectra computed with a low Pb abundance and [Pb/Fe]=+3.25. The HIRES 
spectrum is smoothed using a Gaussian 
having a FWHM of 0.05\AA; this smoothing does not degrade the resolution 
appreciably, while clearly improving the S/N.
\label{fig_pb}}
\end{figure}

\end{document}